\documentclass[10pt,emulateapj,apj]{emulateapj}

\shorttitle{Galaxy Interactions in Clusters}
\shortauthors{Park \& Hwang}
\begin{document}
\title{Interactions of Galaxies in the Galaxy Cluster Environment}
%\twocolumn[
\author{Changbom Park and Ho Seong Hwang}
\affil{School of Physics, Korea Institute for Advanced Study, Seoul 130-722, Korea}
\email{cbp@kias.re.kr, hshwang@kias.re.kr}

\begin{abstract}
We study the dependence of galaxy properties on the clustercentric
  radius and the environment attributed to the nearest neighbor galaxy
  using the Sloan Digital Sky Survey (SDSS) galaxies associated with the Abell galaxy clusters.
We find that there exists a characteristic scale where the properties 
  of galaxies suddenly start to depend on 
  the clustercentric radius at fixed neighbor environment.
The characteristic scale is $1\sim 3$ times the cluster virial radius
  depending on galaxy luminosity.
Existence of the characteristic scale means that the local galaxy
  number density is not directly responsible for the morphology-density relation
  in clusters because the local density varies smoothly with the
  clustercentric radius and has no discontinuity in general. 
What is really working in clusters is the morphology-clustercentric
  radius-neighbor environment relation, where the neighbor environment
  means both neighbor morphology and the local mass density attributed to the neighbor.
The morphology-density relation appears working only because of the statistical correlation 
  between the nearest neighbor distance and the local galaxy number density.
We find strong evidence that
  the hydrodynamic interactions with nearby early-type galaxies 
  is the main drive to quenching star formation activity of late-type galaxies in clusters. 
The hot cluster gas seems to play at most a minor role down to one tenth
  of the cluster virial radius.
We also find that
  the viable mechanisms which can account for the clustercentric radius
  dependence of the structural and internal kinematics parameters are
  harassment and interaction of galaxies with the cluster potential.
The morphology transformation of the late-type galaxies in clusters
  seems to have taken place through both galaxy-galaxy hydrodynamic interactions and 
  galaxy-cluster/galaxy-galaxy gravitational interactions.

\end{abstract}


\section{Introduction}

In galaxy clusters the average morphology of galaxies changes 
with clustercentric radius or local density. The central region of 
clusters at the present epoch is dominated by early-type galaxies. 
This is known as the morphology-radius or morphology-density
  relation (hereafter MRR and MDR, respectively). 
Since the local density is on average a monotonically decreasing function of clustercentric radius,
  they appear to convey us the same information \citep{hh31,dre80,dre97,treu03,smi05,post05,wei06}.
However, there is discussion about which one plays a critical role 
  in determining the galaxy morphology between the clustercentric radius and the local density
  \citep{whi93,dml01,goto03,tk06}.
There is also a remaining issue 
  what physical parameters of galaxies (morphology, color, star formation rate, or stellar mass)
  correlate fundamentally with the environment (e.g., \citealt{cz05,qui06,pog08,ski08}).

A number of physical mechanisms have been proposed to explain this morphology-environment relation.
It is suggested that tidal interactions between individual galaxies 
  can change mass profile and transform
  disk galaxies into spheroidals \citep{bv90}. 
Frequency of galaxy-galaxy interaction depends 
  strongly on the clustercentric radius, and thus galaxy interactions can result in the MRR or MDR.
However, the duration of tidal interactions and thus the total tidal energy deposit 
  are expected to be too small to change galaxy properties significantly 
  for cluster galaxies that are moving at high speeds.
Due to the fast orbital motions the frequency of galaxy mergers is 
  also small in clusters.
A series of frequent high-speed tidal interactions among cluster galaxies,
  called harassment, can produce impulsive heating and cause morphology
  transformation \citep{moo96}.
Tidal interaction with the whole cluster potential well can
also change the structure and activity of galaxies significantly \citep{mw00,gne03}.
%The mechanisms based on tidal interaction are more effective for low
%luminosity galaxies as they are gravitationally more weakly bound.
%But most massive bright galaxies in cluster centers are early types, which seems
%inconsistent with the tidal picture.

Existence of the hot X-ray emitted by intracluster gas makes the
  mechanisms relying on hydrodynamic processes appear attractive.
Stripping of cold gas from infalling late-type galaxies by a ram pressure of the 
  hot intracluster gas, was proposed to explain the increase of the early-type
  fraction toward the cluster center \citep{gg72}.
Hot gas removal from infalling galaxies through hydrodynamic interactions with the 
  intracluster medium can shut off gas supply and star formation activity 
  (SFA) after the already existing cold gas is consumed. 
%This supply-driven
%quenching of SFA, called strangulation, may be effective at cluster periphery.
The interstellar medium of a galaxy traveling in the hot intracluster medium
  can be stripped from the disk due to a viscosity momentum transfer \citep{nul82} or
  evaporate as the temperature of the interstellar medium rises \citep{cs77}.
These hydrodynamic processes have a common drawback that they can
  turn spirals only to S0's and do not produce ellipticals. 
The major effect of the hydrodynamic processes is quenching the SFA by removing or ionizing 
  the cold gas in the disk of late-type galaxies. The bulge-to-disk ratio is not
  expected to increase by these processes.
Therefore, none of the mechanisms proposed so far is able to fully account for
  the MRR or MDR observed in clusters.

Recently, \citet{park08} and \citet{pc09} have found that
  galaxy properties like morphology and luminosity depend strongly on
  the distance and morphology of the nearest neighbor galaxy. 
This dependence was found even when the large-scale background density is fixed, and
  thus is completely different from the commonly known morphology-local density relation. 
Most importantly, the effects of the nearest neighbor change at the
  characteristic scale given by the virial radius of the neighbor galaxy.
When a galaxy is located within the virial radius of its nearest neighbor, 
  its morphology tends to be the same as that of the neighbor. 
But such tendency disappears when the separation is larger than the virial radius.
This fact strongly suggests that galaxies interact hydrodynamically 
  when the separation to their nearest neighbor is smaller than the virial
  radius of the neighbor, and that the effects of such interaction are
  significant enough to change their morphology and SFA.

Outside the virial radius galaxy morphology still depends on
  the distance to the nearest neighbor, but is suddenly independent of 
  neighbor's morphology. The probability for a galaxy to be an early type
  monotonically decreases as the separation increases. 
This strongly supports the idea that the conformity in morphology of galaxy pairs 
  is not primordial but an acquired one through interactions since there should be 
  no reason for the break in morphology conformity to be 
  at the current virial radius of the neighbor if it is initially given.
\citet{park08} proposed that the mechanism responsible for this be
  tidal interactions causing both gravitational and hydrodynamic effects.
Outside the virial radius the interactions are purely gravitational 
  between the galaxy plus dark halo systems,
  but within the virial radius hydrodynamic (and radiative) effects must be involved as well.

A series of tidal interactions and mergers will keep the total cold 
  gas contents in bright galaxies decreasing, producing more early-type
  galaxies as time passes \citep{park08,pc09,hp08}.
The speed of this process is an increasing function of the large-scale density. 
Namely, even though the direct physical processes affecting galaxy morphology are the gravitational and 
  hydrodynamic interactions between neighboring galaxies, the large-scale
  background density appears to control galaxy morphology through its
  statistical correlation with the frequency and strength of galaxy-galaxy interactions.
Given the knowledge that the MDR in most region of the universe 
  (note that Park et al. did not resolve the cluster regions) is the result of galaxy-galaxy
  interactions, one can naturally suspect the MRR and MDR in clusters are also
  due to the interactions between individual galaxies. 
It is the purpose of this paper to explore this possibility.

\section{Observational Data Set}
\subsection{Sloan Digital Sky Survey Sample}

We use a spectroscopic sample of galaxies in the
  Sloan Digital Sky Survey (SDSS) Data Release 6 (DR6; \citealt{ade08}).
The survey produced five-band ({\it ugriz}) photometric
  data for 230 million objects over 8,400 deg$^2$, 
  and optical spectroscopic data more than one million objects of galaxies, quasars, and stars 
  over 6860 deg$^2$ \citep{gunn98,gunn06,uom99,cas01,bla03,fuk96,lup02,hogg01,smi02,ive04,tuc06,pier03}.
Extensive description of SDSS data products is given by \citet{york00} and \citet{sto02}.

The data is supplemented by several value-added galaxy catalogs (VAGCs) drawn from SDSS data.
Photometric and structure parameters of galaxies are obtained from the SDSS pipeline \citep{sto02}.
Complementary photometric parameters such as color gradient, concentration index, and Petrosian radius
  are taken from the DR4plus sample of \citet{choi07}.
The spectroscopic parameters are obtained from
  MPA/JHU and NYU VAGCs \citep{tre04,bla05}.
%In addition, we used the velocity dispersion values of the galaxies
%  given in New York University Value-Added Galaxy Catalog (NYU-VAGC; Blanton et al. 2005).
%that contains the results of a spectral energy distribution fit with sophisticated model

%Figure 1 %%%%%%%%%%%%%%%%%%%%%%%%%%%%%%
\begin{figure}
\center
\includegraphics[scale=0.45]{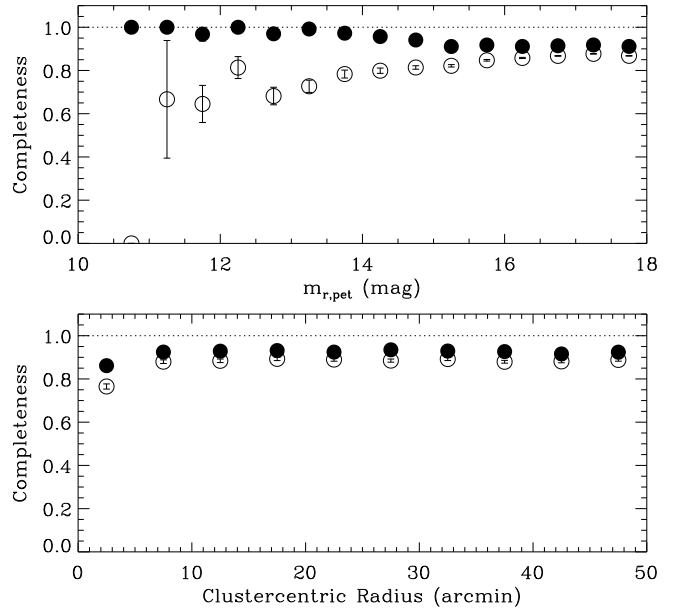}
\caption{Completeness of the spectroscopic sample of cluster galaxies as a function
  of $r$-band magnitude ({\it upper panel}) and clustercentric radius ({\it lower panel}).
Filled circles are the spectroscopic completeness of the data complemented by NED,
  while open circles show the completeness of the original SDSS spectroscopic sample.
}\label{fig-comp}
\end{figure}

Completeness of the spectroscopic data in SDSS is poor 
  for bright galaxies with $m_r<14.5$ because of
  the problems of saturation and cross-talk in the spectrograph, and
  for the galaxies located in high density regions such as galaxy clusters
  due to the fiber collision 
  (two spectroscopic fibers cannot be placed closer than $55\arcsec$ on a given plate).
Therefore, it is necessary to supplement the galaxy data 
  to reduce the possible effects of the incompleteness problem.
We search for the galaxies within ten times the virial radius of each galaxy cluster
  in the photometric catalog of the SDSS galaxies,
  and find their redshifts from the NASA Extragalactic Database (NED)
  to supplement our spectroscopic sample.
Figure \ref{fig-comp} shows the spectroscopic completeness of our galaxy sample
  as a function of apparent magnitude and of clustercentric distance.
The open circles are the completeness of the original SDSS sample,
  and the filled circles are that of our sample with additional redshifts.
It shows that the spectroscopic completeness of our sample
  is higher than $85\%$ at all magnitudes and clustercentric radii.

\subsection{Cluster Sample and Galaxy Membership in Clusters}\label{membership}

We used the Abell catalog of galaxy clusters \citep{aco89} to
  identify cluster galaxies in our galaxy sample.
Among the Abell clusters,
  we selected those that have known spectroscopic
  redshifts in the NED.
We found $730$ clusters located within the SDSS survey region.
We adopted the position of cluster center in the NED,
  but replaced it with the X-ray determined position if it is available in the literature.

In order to determine the membership of galaxies in a cluster,
  we used the ``shifting gapper'' method of \citet{fadda96}
  that was used for the study of kinematics of galaxy clusters \citep{hl07,hl08}.
In the radial velocity versus clustercentric distance space, 
the cluster member galaxies were selected by grouping galaxies
with connection lengths of 950 km s$^{-1}$ in the direction of the radial velocity
and of 0.1 $h^{-1}$Mpc in the direction of the clustercentric radius $R$.
%Grouping in the radial velocity direction is made within each distance bin
%with 0.2 $h^{-1}$Mpc width.
%A larger bin width is used when the number of galaxies in a bin is less than 15.
If the boundary was not reached out to $R=3.5 h^{-1}$Mpc, we stopped
  the grouping at $R=3.5h^{-1}$Mpc.
%  we selected the member galaxies using a
%  velocity gap of 950 km s$^{-1}$ and a distance bin of 0.2 $h^{-1}$ Mpc
%  shifting along the distance from the cluster center.
%We applied this method to the galaxies within the radius
%  at which the distance between adjacent galaxies becomes
%  larger than 0.1 $h^{-1}$ Mpc.
%
We iterated the procedure until the number of cluster members converges.
From this procedure we
  obtained 200 Abell clusters that have more than or equal to 10 member galaxies.

We computed a radius of $r_{200,{\rm cl}}$ (usually called the virial radius)
  for each cluster where the mean overdensity drops to 200
times the critical density of the universe $\rho_{\rm c}$,
  using the formula given by \citet{car97}:
\begin{equation}
r_{200,{\rm cl}}= \frac{3^{1/2}\sigma_{\rm cl}}{10 H(z)},
\end{equation}
where $\sigma_{\rm cl}$ is a velocity dispersion of a cluster and
  the Hubble parameter at $z$ is
  $H^2(z)=H^2_0 [\Omega_m(1+z)^3 +\Omega_k(1+z)^2+\Omega_\Lambda]$ \citep{pee93}.
$\Omega_m$, $\Omega_k$, and $\Omega_\Lambda$ are the dimensionless density parameters.
The velocity dispersion was computed for each cluster
from the redshift distribution of the cluster member galaxies
as described in Appendix.

%Figure 2 %%%%%%%%%%%%%%%%%%%%
\begin{figure}
\center
\plotone{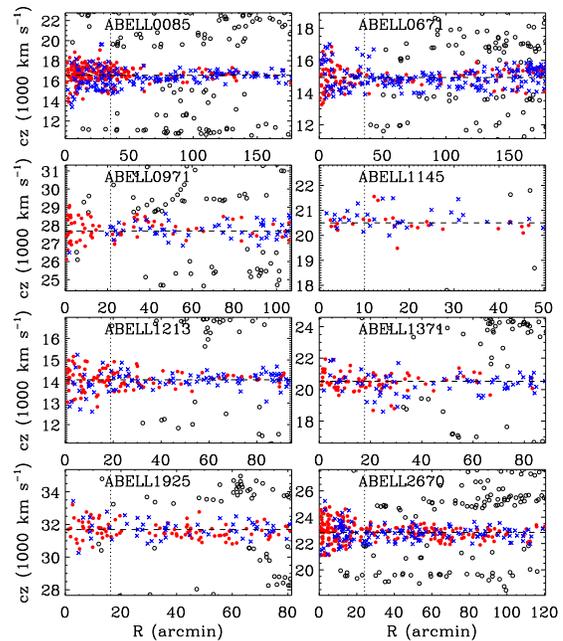}
\caption{Radial velocity vs. clustercentric distance of galaxies.
  Filled circles and crosses indicate the early and late types, respectively, 
  selected as galaxies associated with clusters,
  while open circles indicate the galaxies not selected as associated members.
The horizontal dashed lines indicate the systemic velocity of the clusters 
  determined in this study.
The vertical dotted lines indicate the radius $r_{\rm 200,cl}$ 
  computed in this study.
}\label{fig-member}
\end{figure}

In addition to the sample of cluster member galaxies obtained 
  by the ``shifting gapper'' method above,
  we included the galaxies located at projected separations of $R_{max}<R<10r_{\rm 200,cl}$ 
  to investigate the variation of galaxy properties
  over a wide range of clustercentric radius.
$R_{max}$ is the largest clustercentric distance of the cluster member galaxies determined above.
These additional galaxies were constrained to have velocity difference
  relative to the cluster's systematic velocity
  less than $\Delta v=|v_{\rm gal}-v_{\rm sys}|=1000$ km s$^{-1}$.
The final sample consists of galaxies smoothly distributed 
  from the cluster center to $R=10r_{\rm 200,cl}$ for each cluster.
Figure \ref{fig-member} shows the radial velocities of galaxies
  around eight clusters in our sample as a function of
  clustercentric distance of galaxies.

We rejected the clusters that appeared to be interacting or merging,
  which was decided in the galaxy velocity versus clustercentric distance space.
Dynamically young clusters 
  having the brightest cluster galaxy (BCG) at large clustercentric distance
  ($R_{\rm BCG}>0.18$ $h^{-1}$ Mpc) were rejected too.
We also eliminated the clusters for which survey coverages were not complete 
out to 10$r_{\rm 200,cl}$.
We finally obtained a sample of 93 relaxed Abell clusters and 34,420 associated galaxies
for our analysis.

\subsection{Physical Parameters of Galaxies}
The physical parameters of galaxies that we consider in this study are
  $r$-band absolute Petrosian magnitude ($M_r$),
  morphology, axis ratio,
  $u-r$ color,
  equivalent width of $H\alpha$ emission line,
  $g-i$ color gradient,
  concentration index ($c_{\rm in}$),
  internal velocity dispersion ($\sigma$), and
  Petrosian radius in $i$-band.
Here we give a brief description of these parameters.

%Figure 3 %%%%%%%%%%%%%%%%%%%%
\begin{figure}
\center
\plotone{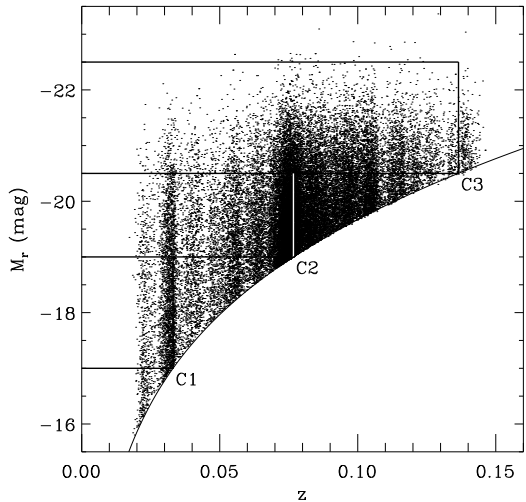}
\caption{Sample definitions of our three volume-limited samples
  in the absolute magnitude vs. redshift space.
  The bottom curve corresponds to the apparent magnitude limit of $m_r=17.77$.
}\label{fig-vol}
\end{figure}

The $r$-band absolute magnitude $M_r$ was computed using the formula,
\begin{equation}\label{eq-mag}
M_r=m_r-5{\rm log}[r(z)(1+z)]-25-K(z)-E(z),
\end{equation}
where $r(z)$ is the comoving distance at redshift $z$ in unit of $h^{-1}$Mpc,
  and the corresponding $5{\rm log}h$ term in $M_r$ will be omitted in this paper.
$K(z)$ is the $K$-correction, and $E(z)$ is the luminosity evolution correction.
We adopt a flat $\Lambda$CDM cosmology with density parameters $\Omega_\Lambda=0.73$ and $\Omega_m=0.27$.
The rest-frame absolute magnitudes of
  individual galaxies are computed in fixed bandpasses, shifted to $z=0.1$,
  using Galactic reddening correction \citep{sch98} and $K$-corrections
  as described by \citet{bla03}.
The evolution correction given by \citet{teg04}, $E(z) = 1.6(z-0.1)$, is also applied.

Figure \ref{fig-vol} shows the $r$-band absolute magnitudes of the cluster galaxies
  against their redshifts.
We define the volume-limited samples of galaxies
  using the redshift and absolute magnitude conditions as follows:
  C1 ($-17.0\geq M_r>-19.0$ and $z\leq0.0331$),
  C2 ($-19.0\geq M_r>-20.5$ and $z\leq0.0767$), and
  C3 ($-20.5\geq M_r>-22.5$ and $z\leq0.1365$).
The apparent magnitude limit line ($m_r=17.77$) shown in Figure \ref{fig-vol} is obtained
  using the mean $K$-correction relation given by equation (2) of \citet{choi07}.

Accurate morphology classification is critical in this work since the effects
  of interaction depend strongly on morphology of the target and neighbor galaxies.
We first classify morphological types of galaxies
  included in the DR4plus sample of \citet{choi07} adopting the automated classification method 
  given by \citet{pc05}. 
Galaxies are divided into early (elliptical and lenticular) and
  late (spiral and irregular) morphological types based on their locations
  in the $u-r$ color versus $g-i$ color gradient space and also in the
  $i$-band concentration index space.
The resulting morphological classification has completeness and reliability reaching 90\%.
The automatic classification scheme does not perform well
  when an early-type galaxy starts to overlap with another galaxy.
This is because the scheme excludes galaxies with very low concentration
  from the early-type class and blended images often erroneously give low concentration.
Since we are investigating the effects of close galaxy-galaxy and galaxy-cluster interactions on galaxy properties,
  this problem in the automatic classification has to be remedied.
We perform an additional visual
  check of the color images of galaxies to correct misclassifications
  by the automated scheme.
In this procedure we changed the types of the
  blended or merging galaxies, blue but elliptical-shaped galaxies,
  and dusty edge-on spirals.
In addition, for the galaxies in DR6 that are not in the DR4plus sample,
  we visually classified their morphological types using the color images.

The $^{0.1}(u-r)$ color was computed using the extinction and $K$-corrected model magnitude.
The superscript 0.1 means the rest-frame magnitude $K$-corrected to the redshift of 0.1,
  and will subsequently be dropped.

We adopt the values of $(g-i)$ color, concentration index ($c_{\rm in}$),
  and Petrosian radius $R_{\rm Pet}$
  computed for the DR4plus sample of galaxies \citep{choi07}.
The $(g-i)$ color gradient was defined by the color difference
  between the region with $R<0.5R_{\rm Pet}$ and the annulus with
  $0.5R_{\rm Pet}<R<R_{\rm Pet}$,
  where $R_{\rm Pet}$ is the Petrosian radius estimated in $i$-band image.
To account for the effect of flattening or inclination of galaxies,
  elliptical annuli were used to calculate the parameters.
The (inverse) concentration index is defined by $R_{50}/R_{90}$,
  where $R_{50}$ and $R_{90}$ are semimajor axis lengths of ellipses
  containing $50\%$ and $90\%$ of the Petrosian flux in the $i$-band image, respectively.

The velocity dispersion value of the galaxy is adopted from NYU-VAGC \citep{bla05}.
The value of $H\alpha$ equivalent width is taken from MPA/JHU-VAGC \citep{tre04},
  which is computed using the straight integration over the fixed bandpass
  from the continuum-subtracted emission line with the model of \citet{bc03}.

In our analysis we often limit the late-type galaxy sample to galaxies with
  $i$-band isophotal axis ratio $b/a$ greater than 0.6.
This is to reduce the effects of internal extinction on our results.
The absolute magnitude and color of late-type galaxies with $b/a < 0.6$ are very inaccurate
  (see Figs. 5 and 12 of \citealt{choi07}).
Therefore, including them in the analysis may introduce
  a large dispersion in SF indicators 
  such as luminosity, color, $H\alpha$ equivalent width, and color gradient.

\subsection{Nearest Neighbor Galaxy in Clusters}

To account for the effects of the nearest neighbor galaxy in cluster environment,
  we determine the distance and the morphology of the nearest neighbor galaxy.

We define the nearest neighbor galaxy of a target galaxy with absolute magnitude $M_r$
  as the one which
    is located closest to the galaxy on the sky and
    is brighter than $M_r+\Delta M_r$ among those in our cluster galaxy sample.
We adopt $\Delta M_r =0.5$.
When we adopt galaxies fainter than the target galaxy by more than 0.5 mag as neighbors,
  our conclusions do not change but our statistics are worse
  since the number of target galaxies becomes smaller (see \S \ref{mer} for more discussion).
%For example, when we adopt $\Delta M_r =0.75$, 
%  the fraction of the target galaxies with $M_r<-20.25$ and $z<0.089$ 
%  whose neighbors are changed 
%  by changing $\Delta M_r$ from 0.5 to 0.75 becomes 28\%. 
%However, our results with $\Delta M_r =0.5$ in the following sections
%  are qualitatively the same with the choice of $\Delta M_r =0.75$ (see also \citealt{park08,pc09}).

We do not use the velocity condition to determine the nearest neighbor galaxy
  because it is selected from the cluster galaxy sample
  to which the velocity condition is already applied.
We obtain the nearest neighbor distance normalized by the virial radius 
  of the nearest neighbor as follows.
We first compute the small-scale density experienced by a target galaxy
  attributed to its neighbor,
\begin{equation}
\rho_{n}/{\bar\rho} = \gamma_n L_n /(4\pi r_p^3 {\bar\rho}/3),
\end{equation}
where $\gamma_n$ is the mass-to-light ratio of the neighbor galaxy,
  $L_n$ is the $r$-band luminosity of the neighbor,
  $r_p$ is the projected separation of the neighbor from the target galaxy, and
  $\bar\rho$ is the mean density of the universe.
We assume that $\gamma$(early)=$2\gamma$(late) at the same $r$-band luminosity,
  and that $\gamma$ is constant with galaxy luminosity for a given morphological type.
The value of mean density of the universe, $\bar\rho=(0.0223\pm0.0005)(\gamma L)_{-20}
(h^{-1}{\rm Mpc})^{-3}$,
  was adopted, where $(\gamma L)_{-20}$ is the mass of a late-type galaxy 
  with $M_r=-20$ \citep{park08}.

Then, we define the virial radius of a neighbor galaxy 
  as the projected radius where the mean mass density $\rho_n$
  within the sphere with radius of $r_p$ is 200 times the critical density
  or 740 times the mean density of the universe, namely,
\begin{equation}
r_{\rm vir} = (3 \gamma L /4\pi / 200{\rho_c})^{1/3}.
\end{equation}
Since we adopt $\Omega_m = 0.27$, $200\rho_c = 200 {\bar\rho}/\Omega_m = 740{\bar\rho}$.
This is almost equal to the virialized density
  $\rho_{\rm virial}=18 \pi^2 / \Omega_m (H_0 t_0)^2 {\bar\rho}= 766{\bar\rho}$
  in the case of our $\Lambda$CDM  universe \citep{gr75}.
This is what \citet{park08} used to define the virial radius.
According to our formula the virial radii of galaxies with
  $M_r=-19.5,-20.0,$ and $-20.5$ are 260, 300, and 350 $h^{-1}$ kpc for early types,
  and 210, 240, and 280 $h^{-1}$ kpc for late types, respectively.

\section{Results}

\subsection{Morphology-Environment Relation}\label{mer}

%Figure 4 %%%%%%%%%%%%%%%%%%%%%%%%%%%%%%
\begin{figure} 
\center
\plotone{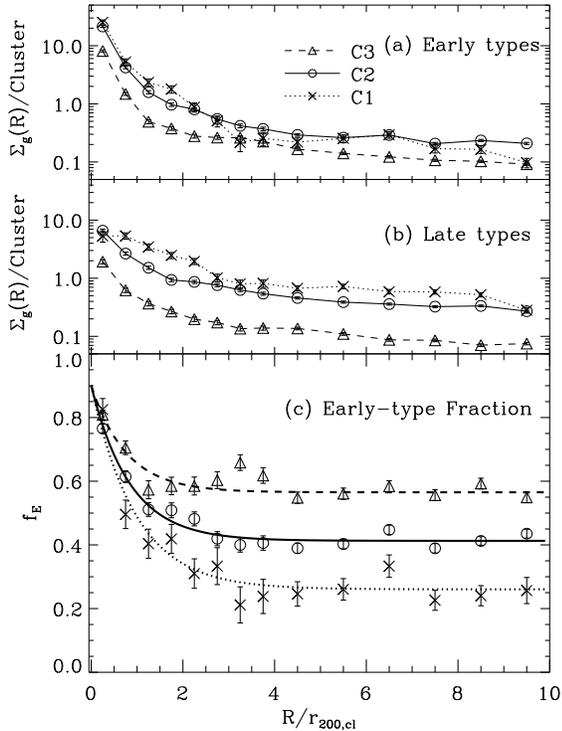}
\caption{
The surface number density of early-type (upper panel) and late-type
(middle panel) galaxies in the Abell cluster regions, and the
corresponding early-type fraction (bottom panel) as a function  
of the clustercentric radius normalized to the cluster virial radius.
The lines in the bottom panel are the best-fitting functions.
}\label{fig-frac1d}
\end{figure}

%Figure 5 %%%%%%%%%%%%%%%%%%%%%%%%%%%%%%
\begin{figure} 
\plotone{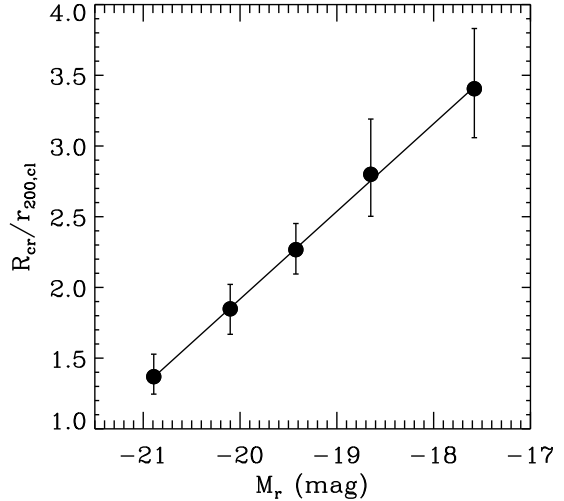}
\caption{
The characteristic clustercentric radius $R_{\rm cr}$ as a function of 
$r$-band absolute magnitude. $R_{\rm cr}$ is the scale where the early-type
fraction $f_E$ starts to rise significantly above the field value.
Five luminosity subsamples are used: $-17\ge M_r > -18, -18\ge M_r>-19,
-19\ge M_r >-19.75, -19.75\ge M_r >-20.5, -20.5\ge M_r >-22.5$.
The best-fit linear line is $R_{\rm cr}/r_{\rm 200,cl} = (0.62\pm 0.10)M_r + 
(14.32 \pm 1.95)$.
}\label{fig-fracdd}
\end{figure}

%Figure 6 %%%%%%%%%%%%%%%%%%%%%%%%%%%%%%
\begin{figure*} 
\plotone{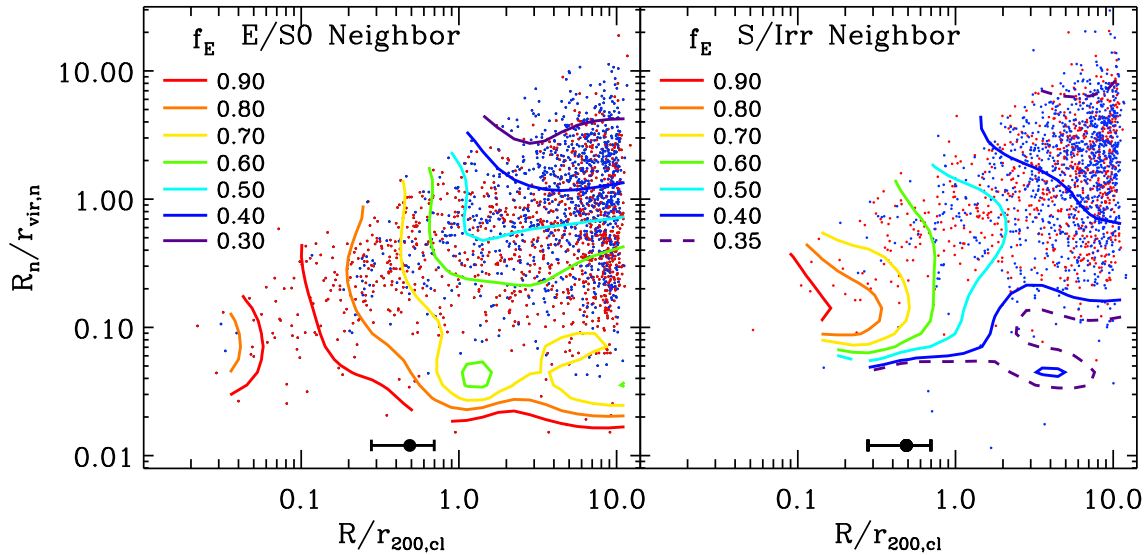}
\caption{
Morphology-environment relation when the nearest galaxy is
  (left) an early-type and (right) a late-type.
Absolute magnitude of galaxies is fixed to a narrow range of $-19.5\ge M_r>-20.0$.
Contours show constant early-type galaxy fraction $f_E$.
The points with error bars above the x-axes denote the average 
  virial radius of the BCGs.
}\label{fig-frac2d}
\end{figure*}

Figure \ref{fig-frac1d} shows the surface number density of galaxies and 
  the fraction of early-type galaxies as a function
  of projected clustercentric radius $R$ normalized by the cluster virial radius $r_{\rm 200,cl}$.
They are shown for three luminosity ranges.
The uncertainties of the fraction represent $68\%$ $(1\sigma)$ confidence intervals 
  that are determined from the numerical bootstrap procedure.

It shows the well-known MRR or MDR of galaxies in clusters. 
In all three luminosity ranges the early-type fraction $f_E$
  is almost constant at large distances, and starts to
  increase inwards at the critical region with $R\approx 1\sim 3 r_{\rm 200,cl}$,
  which corresponds to the cluster infall region.
The transition radius seems depending on galaxy luminosity.
For relatively brighter galaxies with $-20.5 \ge M_r > -22.5$ (the subsample C3)
  the transition occurs at $\sim1.5r_{\rm 200,cl}$, but
  for fainter galaxies with $-17.0\ge M_r > -19.0$ (the subsample C1)
  it occurs at $\sim3 r_{\rm 200,cl}$. 
The slope of increase is larger for less luminous galaxies at small $R$.
Since we are not distinguishing between ellipticals and lenticulars, 
  we can only see their sum is rising toward the cluster center.
Fainter galaxies seem more vulnerable to the cluster influence.
This can be considered as evidence for the direct interaction between
  clusters and infalling galaxies, thinking that less massive galaxies 
  change their morphology at farther distances from clusters relative to massive ones. 
However, it can be also due to the mass segregation within clusters that 
  massive cluster member galaxies hardly overshoot beyond the cluster virial radius 
  while the less massive ones overshoot out to a few times the virial radius. 
Massive galaxies outside the virial radius are likely to be the infalling ones 
  rather than bound cluster members.

Figure \ref{fig-frac1d} indicates that the early-type fraction is nearly the same
  near the cluster center regardless of galaxy luminosity. 
The central value of $f_E=0.8\sim 0.85$ actually means that the fraction
  is close to 1.0 at the cluster center if the projection effect is taken
  into account \citep{ann08}.
At radii larger than about $4 r_{\rm 200,cl}$, $f_E$ approaches
  the field fraction that depends on the luminosity range 
  and becomes insensitive to the large-scale environment \citep{park07}.

To inspect this characteristic scale in more details we fit $f_E$ as a function
of $R/r_{\rm 200,cl}$ using the function
\begin{equation}
f_E = (0.9-f_0){\rm exp}(-R/R_0)+f_0,
\end{equation}
where $R_0$ is a free parameter and $f_0$ is an average value of $f_E$ at 
  $R/r_{\rm 200,cl} > 4.0$. 
We define the characteristic radius $R_{\rm cr}$ as the scale
  where $f_E$ becomes 10\% larger than its field value at fixed luminosity.
Figure \ref{fig-fracdd} shows an almost linear relation between the characteristic radius and
  the absolute magnitude obtained from five luminosity subsets.
The best-fit linear line is $R_{\rm cr}/r_{\rm 200,cl} = (0.62\pm 0.10)M_r + (14.32 \pm 1.95)$.

We now investigate the dependence of galaxy morphology
on both clustercentric radius and nearest neighbor distance. 
We select a volume-limited sample of target galaxies associated with clusters 
  whose absolute magnitudes are in a narrow range of $-19.5\geq M_r>-20.0$
  and redshifts are less than $0.0767$.
Their nearest neighbors are found among all galaxies with $M_r\leq-19.0$.
Then $r_{\rm vir}$ of the nearest neighbors are calculated using 
  their luminosity and morphology.

Dots in Figure \ref{fig-frac2d} show the distribution of early-type ({\it red points})
  and late-type ({\it blue points}) galaxies in the projected 
  clustercentric radius $R$ and projected nearest neighbor distance $R_n$ space.
A spline kernel is used to obtain the smooth distributions of the median $f_E$ in 
  each location of the two panels. 
Contours with different colors mark constant early-type fractions.
The left panel of Figure \ref{fig-frac2d} contains the target galaxies having early-type neighbors. 
It shows that at $R \ga 2 r_{\rm 200,cl}$ all contours are nearly horizontal.
This means that outside the cluster virial radius
  galaxy morphology is determined solely by 
  the nearest neighbor distance and morphology.
When the nearest neighbor is an early type, 
  $f_E$ monotonically increases as $R_n$ decreases. 
But if the neighbor is a late type, 
  it first increases, reaches a maximum at $R_n \sim r_{\rm vir,n}/3$, 
  and then decreases as $R_n$ decreases. 
The bifurcation occurs at $R_n \sim r_{\rm vir,n}$.
This is the morphology-neighbor environment relation 
  that was discovered by \citet{park08} in the general large-scale
  background density environment. 
It is interesting to see that the effects of the nearest neighbor's distance and morphology 
  are the dominant factors of galaxy morphology transformation right down to 
  the cluster infall regions.

The situation abruptly changes at the critical clustercentric radii of
  $R_{\rm cr}=1\sim3 r_{\rm 200,cl}$. 
Within this radius contours are nearly vertical 
  when $R_n \ga 0.1r_{\rm vir,n}$,
but are nearly horizontal when $R_n\la 0.05 r_{\rm vir,n}$.
When $R_n \ga 0.1 r_{\rm vir,n}$, 
  $f_E$ monotonically increase as $R$ decreases and 
  depends almost entirely on $R$ regardless of the morphological
  type of the nearest neighbor. 
Existence of this sudden transition near the cluster virial radius
  suggests that the morphology transformation of cluster galaxies is due to
  the interactions between galaxies and clusters.
Both gravitational and hydrodynamic processes can produce the discontinuity in $f_E$.
It may be the hot cluster gas confined within $\sim r_{\rm 200,cl}$ or
  gravitational tidal force of cluster acting on galaxies trapped within
  the virial radius that causes the transformation.
The monotonic increase of $f_E$ at smaller $R$ can arise
  by the increase of the effects of hot cluster gas whose pressure
  monotonically increases toward the cluster center and/or
  by interactions with the cluster potential for galaxies 
  with smaller orbital radii that have the shorter crossing time.

On the other hand, Figure \ref{fig-frac2d} clearly indicates that the clustercentric 
  radius is not the only environmental parameter determining the galaxy
  morphology but the nearest neighbor does a critical role
  when the neighbor separation is less than about $0.05 r_{\rm vir,n}$.
The local galaxy number density, to which the MDR is often attributed, 
  cannot be responsible for the increase of $f_E$ inwards cluster
  because the density rises smoothly as $R$ decreases and does not have a
  characteristic break at $R_{\rm cr}$ (see Fig. \ref{fig-frac1d}).
After all, it is the morphology-clustercentric radius-neighbor environment 
  relation instead of the simple MDR or MRR that determines the morphology of cluster galaxies.
Here neighbor environment includes both neighbor distance
  and neighbor morphology.

We emphasize that we are not using the nearest neighbor distance
as a measure of local galaxy number density. It is a measure of
the influence of the nearest neighbor galaxy itself irrespectively 
of other galaxies. However, there exists statistical correlation between
$R_n$ and the local galaxy number density.
If one uses a measure of local galaxy number density, one will still
  find some correlation of the measure with galaxy morphology 
  within the cluster virial radius. 
But it is actually the galaxy-galaxy interaction that causes the correlation.
This is because fixing $R$ essentially fixes the local galaxy number density
  and in Figure \ref{fig-frac2d} $f_E$ still changes at a fixed $R$.

In the left panel of Figure \ref{fig-frac2d} galaxies have early-type neighbors.
It is expected that both cluster and neighbor galaxy impose morphology transformation 
  toward early type on galaxies through hydrodynamic or gravitational effects. 
Dominance of the role between the two depends on $R$ and $R_n$.
At $R\approx r_{\rm 200,cl}$
  the neighbor galaxy starts to give a major impact on
  galaxy morphology at the separations $R_n < 0.5 r_{\rm vir,n} = 100\sim 150 h^{-1}$kpc 
  (note the difference of $f_E$ between the two panels).
But at $R\approx 0.2 r_{\rm 200,cl}$ it happens at 
  $R_n < 0.05 r_{\rm vir,n}\approx 10\sim15 h^{-1}$kpc.
This seems reasonable since near the cluster center
  the tidal effects by the cluster potential are stronger,
  the cluster gas has higher pressure,
  and thus the cluster gives more direct impact on galaxy properties.

One can note from a comparison between two panels of Figure \ref{fig-frac2d} that
  directions and levels of contours 
  change completely depending on the morphology of the nearest
  neighbor galaxy at any clustercentric radius when the pair separation
  is small (i.e. $R_n \la 0.1 r_{\rm vir,n}$). 
The right panel shows that a cluster galaxy tends to become
a late type when it has a late-type neighbor within the separation
$R_n \approx 0.1 r_{\rm vir,n}$ even when the pair is well 
 within the cluster virial radius. 
For pairs with $R_n \approx 0.05 r_{\rm vir,n}$ 
  $f_E\approx 0.8$ at the clustercentric radius of $R=0.5 r_{\rm 200,cl}$
  when the neighbor is an early type. 
But $f_E$ is as low as 0.35 for pairs with the same $R_n$ located
at the same $R$ when the neighbor is a late type.
It is a clear demonstration that the nearest neighbor has a dominant
  control over galaxy morphology transformation in clusters
  when the distance to the neighbor is less than about one tenth of the
  virial radius of the neighbor galaxy. 

Figure \ref{fig-frac2d} shows that there are not many such short separation
  pairs at a particular moment
  within the cluster virial radius and one might think that the majority of cluster galaxies
  are not affected by neighbors.
However, it can be seen in Figure \ref{fig-frac2d} that at $R\la 0.5r_{\rm 200,cl}$ 
  the mean separation between galaxies becomes so small
  that the virialized regions associated with galaxies are all overlapping with one another
  (not that the upper edge of the galaxy distribution becomes smaller than 
  $R_n /r_{\rm vir,n}=1$ at $R/r_{\rm 200,cl}\la0.5$ in Fig. \ref{fig-frac2d}). 
Suppose the RMS orbital velocity of galaxies within $r_{\rm 200,cl}$ is $\langle v^2 \rangle^{1/2}$.
If $r_{\rm 200,cl}=1.5$ $h^{-1}$ Mpc and $\langle v^2\rangle^{1/2}=700$ km s$^{-1}$,
  the crossing time across $r_{\rm 200,cl}$ is about $3\times10^9$ yrs.
On the other hand, 
  the RMS relative velocity between two random cluster galaxies is $\sqrt2\langle v^2 \rangle^{1/2}$.
If the virial radius of the galaxies is $r_{\rm vir}=300$ $h^{-1}$ kpc,
  their crossing time is $4\times10^8$ yrs, 
  an order of magnitude shorter than the crossing time across the cluster.
This means in this figure that the cluster galaxies at $R<r_{\rm 200,cl}$ have migrated vertically 
  (interaction with neighbor galaxies) many times
   during the time they make one travel horizontally (interaction with the cluster).
In particular, when $R<0.5r_{\rm 200,cl}$,
  a galaxy starts to interact with another neighbor galaxy
  as it passes one neighbor galaxy 
  because the virial radii of galaxies all overlap with one another there.
Then the nearest neighbors may have left significant cumulative effects
  on morphology of cluster galaxies. 
The amount of the cumulative effects
  increases monotonically as the clustercentric radius decreases, which can produce
  the trend seen in Figure \ref{fig-frac2d}.

It is expected that the late-type galaxy sample has a higher fraction of interlopers 
  in the cluster environment.
The pairs involving late-type target or late-type neighbor are more likely to be 
  false pairs, and more widely separated in real space. 
If this effect is corrected in Figure \ref{fig-frac2d}, 
  the contours will be shifted upward. 
Furthermore, if the fraction of interlopers varies as a function of neighbor
  separation or clustercentric radius, the contours will be both shifted and stretched.
However, the fact that there are discontinuities in $f_E$ shown in Figure \ref{fig-frac2d} 
  (near the cluster virial radius $R \approx r_{\rm 200,cl}$ 
  and the galaxy merger scale $R_n \approx 0.05 r_{\rm vir,n}$) 
  and also the fact that these transitions occur at physically meaningful scales, 
  cannot be explained if the fraction of interlopers is seriously high.
This suggests that the morphology-environment relation shown by the contours 
  in Figure \ref{fig-frac2d} is not seriously affected by the interlopers.

Though we rejected the clusters that appear to be interacting or merging,
  there are some clusters ($\sim$15\%) whose projected distance to the nearest cluster
  is less than 10$r_{\rm 200,cl}$ and relative velocity is less than 1000 km s$^{-1}$.
Therefore, the galaxies associated with these clusters
  have been counted more than once in our analysis.
For example, among 3874 galaxies used in Figure \ref{fig-frac2d},
  874 galaxies are those counted more than once.
When we reject these galaxies, our results remain almost the same.
In addition, our results do not depend sensitively on our definition of the nearest neighbor.
For example, when we varied $\Delta M_r$ from 0.5 to 0.75,
  our results remain qualitatively the same (see also \citealt{park08,pc09}).

The mean virial radius of the BCGs, which is typically about $0.5 r_{\rm 200,cl}$, 
  and its $1 \sigma$ range are shown above the $x$-axes of Figure \ref{fig-frac2d}. 
We do not find any characteristic feature across the virial radius of the BCGs. 
This suggests that the BCGs are not directly participating in transforming
  the morphology of the cluster galaxies.
When the positions of BCGs are used as the centers of clusters,
  the contours in Figure \ref{fig-frac2d} are basically unchanged at $R\ga 0.2r_{\rm 200,cl}$.
But the slight drop of $f_E$ at $R< 0.1r_{\rm 200,cl}$ seen for the E/S0 neighbor 
  case, now disappears.

%Figure 7 %%%%%%%%%%%%%%%%%%%%%%%%%%%%%%
\begin{figure} 
\center
%\plotone{f6.eps}
\includegraphics[scale=0.7]{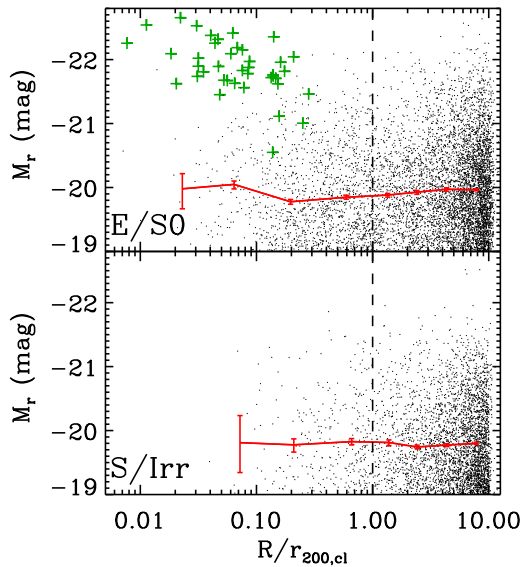}
\caption{
Absolute magnitude of galaxies brighter than $M_r=-19$ 
 in the Abell cluster regions
as a function of clustercentric radius. The upper panel shows the early types,
and the lower panel shows the late types. Lines are the median magnitudes.
Crosses are the BCGs, and they are not used in calculating the median values.
Late types with axis ratio of $b/a<0.6$ were eliminated.
The vertical dashed lines indicate the cluster virial radius $r_{\rm 200,cl}$.
}\label{fig-mag1d}
\end{figure}

%Figure 8 %%%%%%%%%%%%%%%%%%%%%%%%%%%%%%
\begin{figure*}
\center
%\plotone{f7.eps}
\includegraphics[scale=0.6]{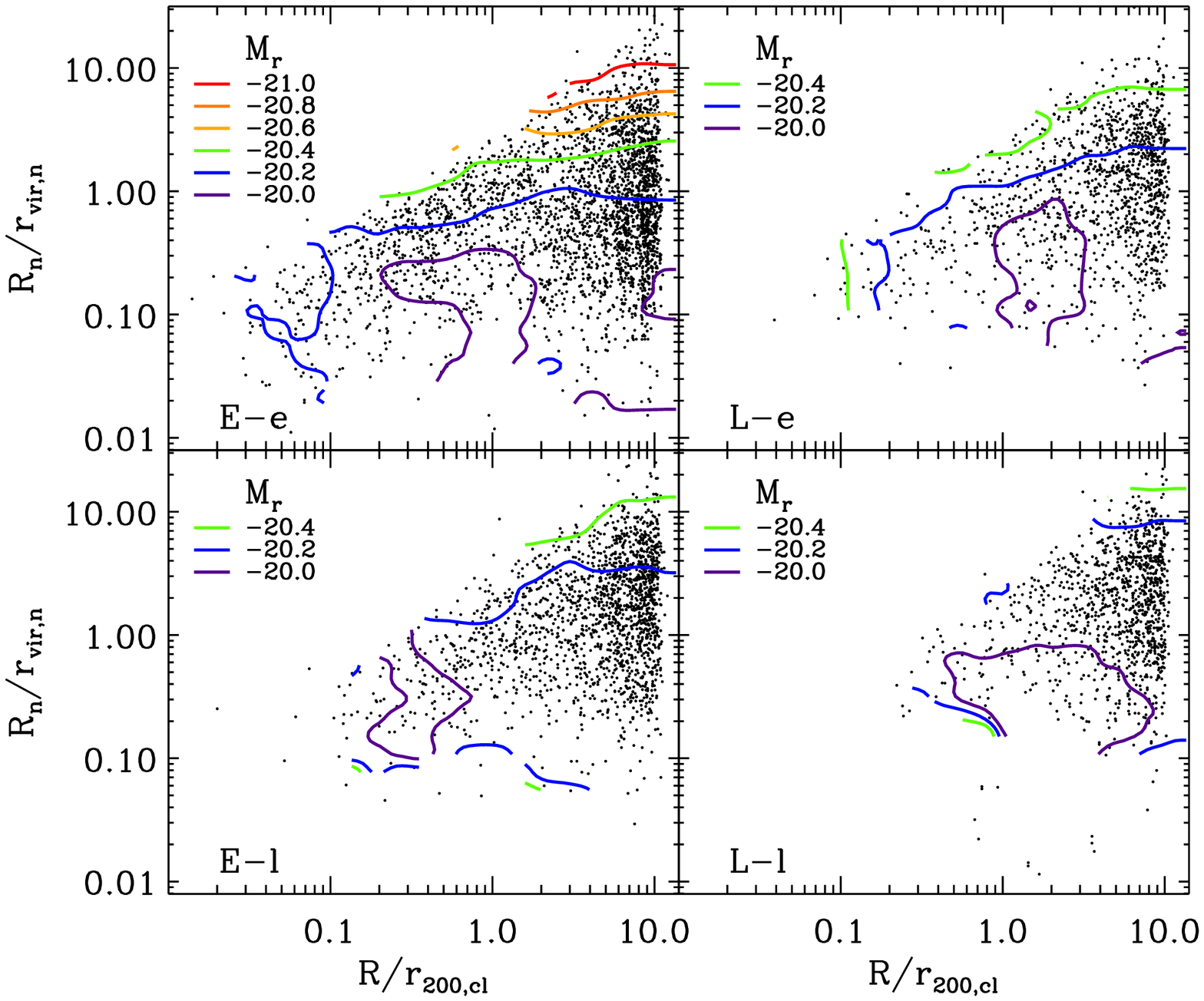}
\caption{
Median absolute magnitude contours in the projected pair separation $R_n/r_{\rm vir,n}$
  vs. the clustercentric distance $R/r_{\rm 200,cl}$ for the galaxies with 
  $z\leq0.0767$ and $M_r\leq-19.5$.  
Late types with axis ratio of b/a$<0.6$ were eliminated.
  Four cases are given;
  the early-type target galaxies having an early-type neighbor (E-e),
  the early-type target galaxies having a late-type neighbor (E-l),
  the late-type target galaxies having an early-type neighbor (L-e), and
  the late-type target galaxies having a late-type neighbor (L-l).
  }\label{fig-mag2d}
\end{figure*}

\subsection{Luminosity-Environment Relation}

Figure \ref{fig-mag1d} shows the $r$-band absolute magnitudes of the galaxies in a 
  volume-limited sample with $z < 0.0767$ and $M_r < -19.0$. 
The lines with error bars are the median values as a function of $R$. 
The BCGs, marked as crosses, are not used in calculating the median values.
Early-type galaxies become fainter as much as $\sim0.2$ magnitude
  as $R$ decreases from $10r_{\rm 200,cl}$ to $0.2r_{\rm 200,cl}$. 
The luminosity of early types rises again as $R$ becomes smaller 
  than about $0.2r_{\rm 200,cl}$.
But the increase of luminosity near the cluster center seems to
  depend on how much the BCGs are off-centered.
Such a variation of the median luminosity is not detected for late types.

%If we reject the clusters whose BCGs are slightly off-centered ($R_{\rm BCGs}>0.1r_{\rm 200,cl}$),
%  the results do not change.
%On the other hand, if we plot Figure \ref{fig-mag1d} using 
%  the only clusters whose BCGs are slightly off-centered,
%  the luminosity of early types rises again at $R\sim0.3-0.4r_{\rm 200,cl}$,
%  which is slightly larger than the case using all the galaxies.
%It suggests that the clusters with off-centered BCGs show the change of galaxy properties
%  at slightly larger clustercentric radius in the central region of the cluster,
%  though we do not know the reason why some BCGs are off-centered
%   (the center of cluster may be not well determined though we tried to supplement the cluster center coordinates with X-ray data,
%   and/or these clusters may be dynamically young).
%However, the off-centered clusters occupy a small fraction in total sample ($<30\%$),
%  the conclusion is not affected by these clusters.

The dependence of luminosity on the environment can be better understood in the two dimensional study. 
Contours in Figure \ref{fig-mag2d} delineate the constant median $M_r$ locations 
  in the two-dimensional environmental parameter space of $R$ and $R_n$.
Four panels distinguish among four different combinations of target and neighbor morphology. 
At each location of the $R$-$R_n$ space the median value of $M_r$ is found
  using the galaxies included within a spline kernel with a fixed size.
The BCGs are not included in the smoothing.

It can be clearly seen that galaxies with larger $R_n$ are brighter 
  when $R$ is greater than about $0.2r_{\rm 200,cl}$. 
Namely, the bright galaxies are more isolated from their influential neighbors than
  the relatively faint galaxies are.
In the region far outside the cluster virial radius the
  absolute magnitude of galaxies is nearly constant in $R$
  (i.e. the contours are nearly horizontal).
Luminosity of galaxies with $R_n\ga r_{\rm vir,n}$ depends very sensitively on $R_n$.
But galaxies with $R_n< r_{\rm vir,n}$ show little dependence of $M_r$ on $R$ or $R_n$.
The same phenomenon has been found by \citet{park08} for 
  galaxy pairs in the general background density regions. 

The reason that galaxies are brighter when they are located outside
  the nearest neighbor's virial radius was interpreted by Park et al. as due to 
  luminosity transformation through mergers. 
After a galaxy merges with
  its closest neighbor, the second nearest neighbor will become the
  new nearest neighbor of the merger product. 
Since the merger product is typically located at a
  separation from the originally second nearest neighbor larger
  than the virial radius of the neighbor, a recently merged galaxy will
  appear in general at the upper edge of the galaxy distribution in Figure \ref{fig-mag2d}.
As one moves toward the cluster center, the average
  galaxy number density increases and the mean distance between galaxies decreases.
Correspondingly, the distance to the new nearest neighbor of a
  recently merged galaxy also decreases statistically for decreasing
  clustercentric radius, and the location of the recent merger products
  follows the upper edge of the galaxy distribution in Figure \ref{fig-mag2d}. 

%Figure 9 %%%%%%%%%%%%%%%%%%%%%%%%%%%%%%
\begin{figure*}
\center
%\plotone{f9.eps}
\includegraphics[scale=0.7]{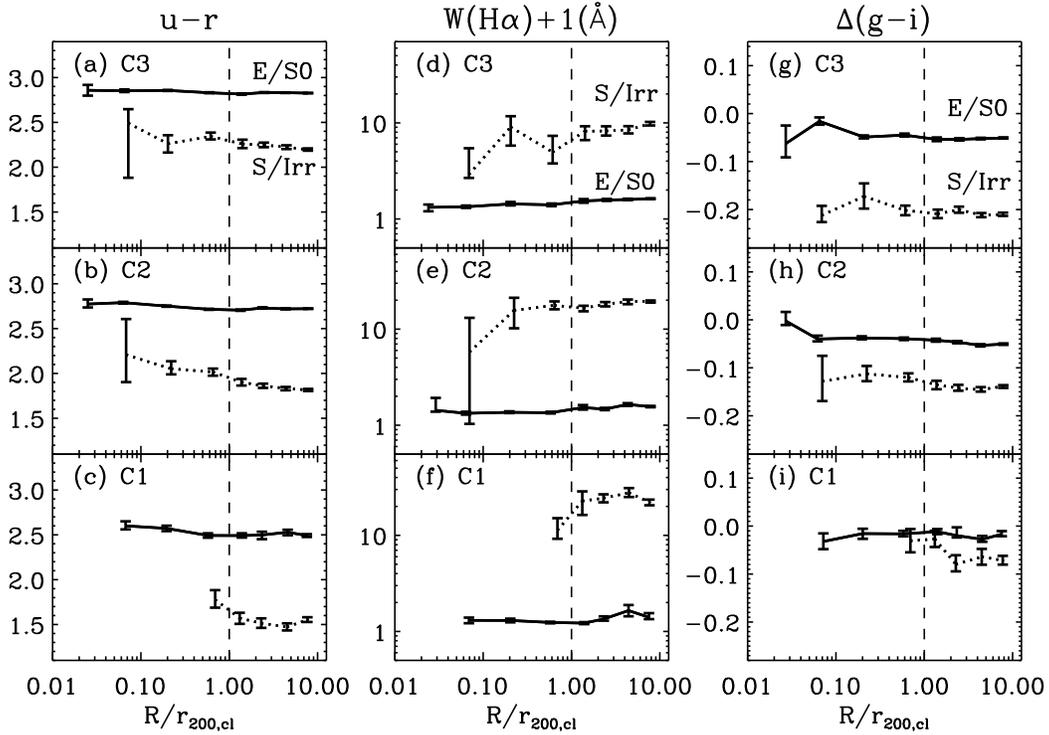}
\caption{
Dependence of star formation activity parameters of our target galaxies in the samples of C1, C2, and C3
   on the clustercentric distance:
  (left) $u-r$, (middle) W(H$\alpha$)+1($\AA$), and (right) $\Delta(g-i)$.
Median curves are drawn for the cases of early types (solid line)
  and of late types (dotted line).
Late types with axis ratio of $b/a<0.6$ were eliminated.}\label{fig-sf1d}
\end{figure*}

%Figure 10 %%%%%%%%%%%%%%%%%%%%%%%%%%%%%%
\begin{figure*}
\center
\plotone{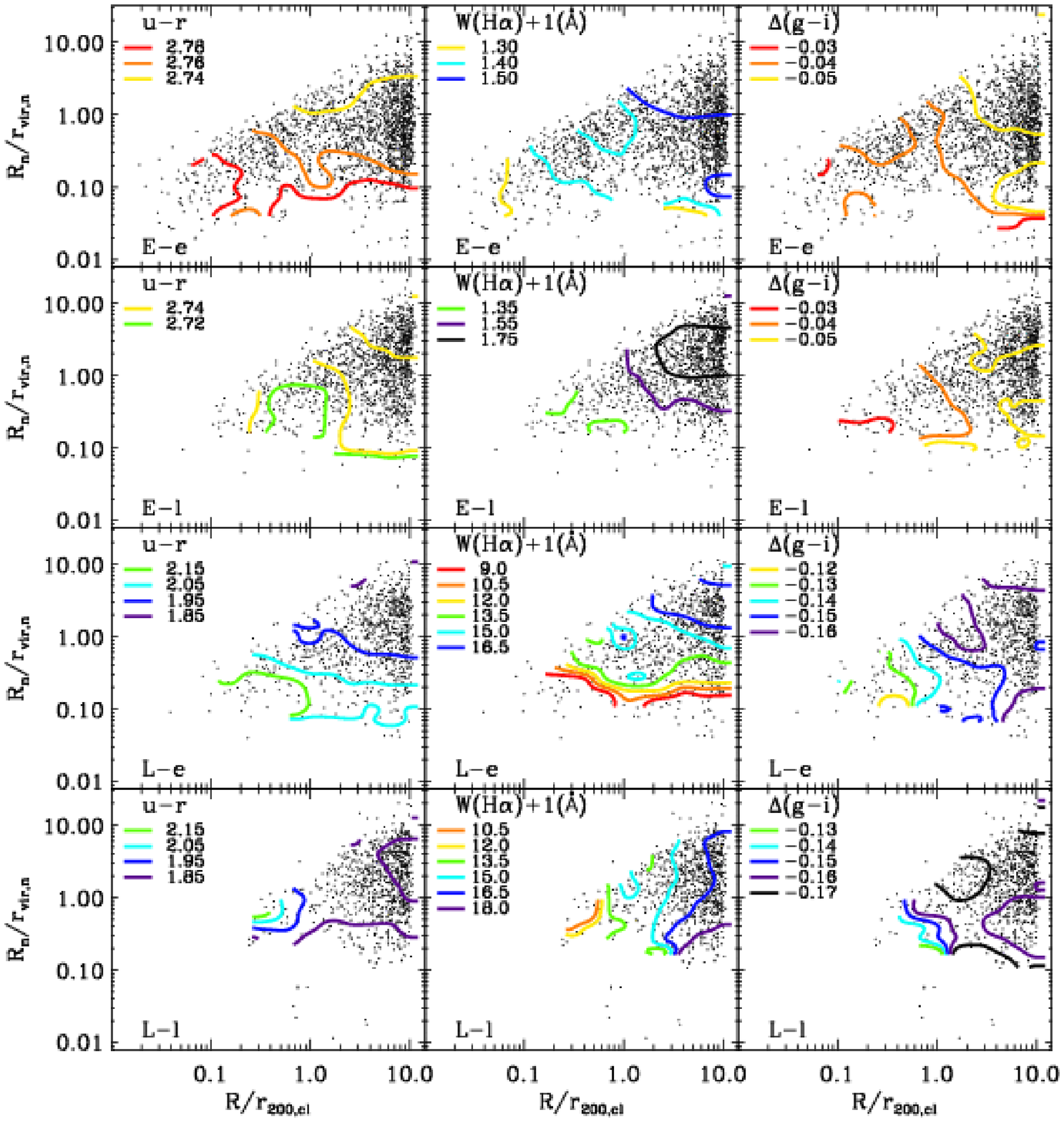}
\caption{
Dependence of $u-r$ color, equivalent width of the $H{\alpha}$ line,
  $g-i$ color gradient of galaxies with $-19.5\ge M_r>-20.5$ on the pair separation
  $R_n$ and the clustercentric distance $R$.
In each column, galaxies are divided into four cases, the E-e, E-l, L-e, and L-l galaxies.
Dots are galaxies belonging to each subset.
At each location of the $R_n/r_{\rm vir,n}$-$R/r_{\rm 200,cl}$ space the median value of the
  physical parameter is found from those of galaxies within a certain
  distance from the location.
Curves are the constant-parameter contours.
Late types with axis ratio of $b/a<0.6$ were eliminated.
}\label{fig-sf2d}
\end{figure*}

The upper left panel of Figure \ref{fig-mag2d}, showing the early-type target galaxies
  having early-type neighbors (the E-e case), shows that the maximum
  luminosity a galaxy can attain slowly decreases as $R$ decreases
  until $R$ becomes about $0.2r_{\rm 200,cl}$.
The median $M_r$ of the most separated galaxies is about $-21.3$
at $R\approx 10 r_{\rm 200,cl}$, but decreases to about $-20.0$ at
$R \approx 0.2 r_{\rm 200,cl}$, a more than one magnitude drop.
This can be explained if the luminosity transformation process
  by galaxy-galaxy mergers becomes less efficient as $R$ decreases.
This is reasonable because the relative velocity between neighboring
  galaxies will be higher and the chance for a pair to merge will 
  decrease toward the cluster center.

A similar $R$-dependent trend in luminosity is also seen 
  for late-type galaxies as shown in the right panels of Figure \ref{fig-mag2d}
  even though the trend is much weaker.
Luminosity of late-type galaxies is on average lower than that of early types.
Interestingly, the late-type galaxies having an early-type
  neighbor (the L-e galaxies) are slightly brighter than the early-type galaxies 
  at $R\la 0.1 r_{\rm 200,cl}$.
It could be that, as $R$ decreases, faint late types are
  transformed to early types and only relatively bright late types
  survive near the cluster center. 
%In particular,
%  this can cancel the trend of decreasing luminosity of late types
%  with decreasing $R$ due to paucity of merger products near the center,
%  and make the typical luminosity of galaxies increase at $R<0.1r_{\rm 200,cl}$.

%The first and third panels of Figure @@ tells that at $R< r_{\rm 200,cl}$ 
%galaxies with early-type neighbor in general become brighter 
%as they approach the cluster center. On the other hand, luminosity 
%of galaxies having late-type neighbors is nearly independent 
%of $R$. Since the cases are distinguished only by neighbor's
%morphology, the difference should be attributed to the neighbor
%and the clustercentric radius dependence of luminosity must
%have been also caused by the neighbor galaxy whose average
%properties depend on $R$. Namely, clusters do only an indirect
%role in transforming galaxy luminosity within their virial radii.

Figure \ref{fig-mag2d} also shows that galaxies located near $R=r_{\rm 200,cl}$
  and having a close neighbor are on average the faintest.
Namely, even though the most isolated galaxies become faintest at $R\approx0.2r_{\rm 200,cl}$,
  those with a close neighbor ($R_n\approx0.1r_{\rm vir,n}$) become faintest
  at $R\approx1.0r_{\rm 200,cl}$.

\subsection{Star Formation Activity Parameters}

The left column of Figure \ref{fig-sf1d} shows the $u-r$ color of galaxies divided into two
  morphology and three magnitude bins as a function of $R$.
It can be seen that the galaxy color becomes redder as $R$ decreases
  for both early and late types.
This clustercentric radius dependence of color is stronger 
  for relatively fainter galaxies (i.e. the C1 sample).
In the case of the galaxies in C1 the dependence seems to occur at $\sim3r_{\rm 200,cl}$.

Dots in Figure \ref{fig-sf2d} are the intermediate luminosity galaxies 
  with $-19.5\ge M_r>-20.5$ in the two-dimensional environmental parameter space.
Color of early-type galaxies is nearly constant everywhere.
There is a slight tendency that early-type galaxies become
  redder as they approach an early-type neighbor.

The constant color contours of late-type galaxies are nearly
  horizontal at $R> R_{\rm cr} = 1\sim 3 r_{\rm 200,cl}$, which means galaxy 
  color is independent of $R$ and depends only on the neighbor separation $R_n$.
Inside the critical radius $R_{\rm cr}$ the color of late-type galaxies still
  depends mainly on neighbor separation and morphology.
%The transition is abrupt particularly for the L-l case.
%Such an abrupt clustercentric radius dependence was also seen
%for morphology and luminosity.
%When $R_n$ is less than about $0.2 r_{200,nei}$, it is the 
%neighbor's morphology and separation that control the color
%of late-type galaxies. 
This is evidence that hydrodynamic
  interactions between individual galaxies are effective even within
  the virialized region of clusters.
In the case of the L-l galaxies there is clustercentric radius dependence of color
  when $R_n\ga r_{\rm vir,n}/3$.
It is likely that this is because the neighbors other than the nearest one 
  tend to be early types as the system moves toward the cluster center.
If the cluster hot gas had a direct impact on late-type galaxies' color,
  both L-e and L-l galaxies would show a similar color dependence on $R$.

%Figure 11 %%%%%%%%%%%%%%%%%%%%%%%%%%%%%%
\begin{figure*}
\center
%\plotone{f11.eps}
\includegraphics[scale=0.7]{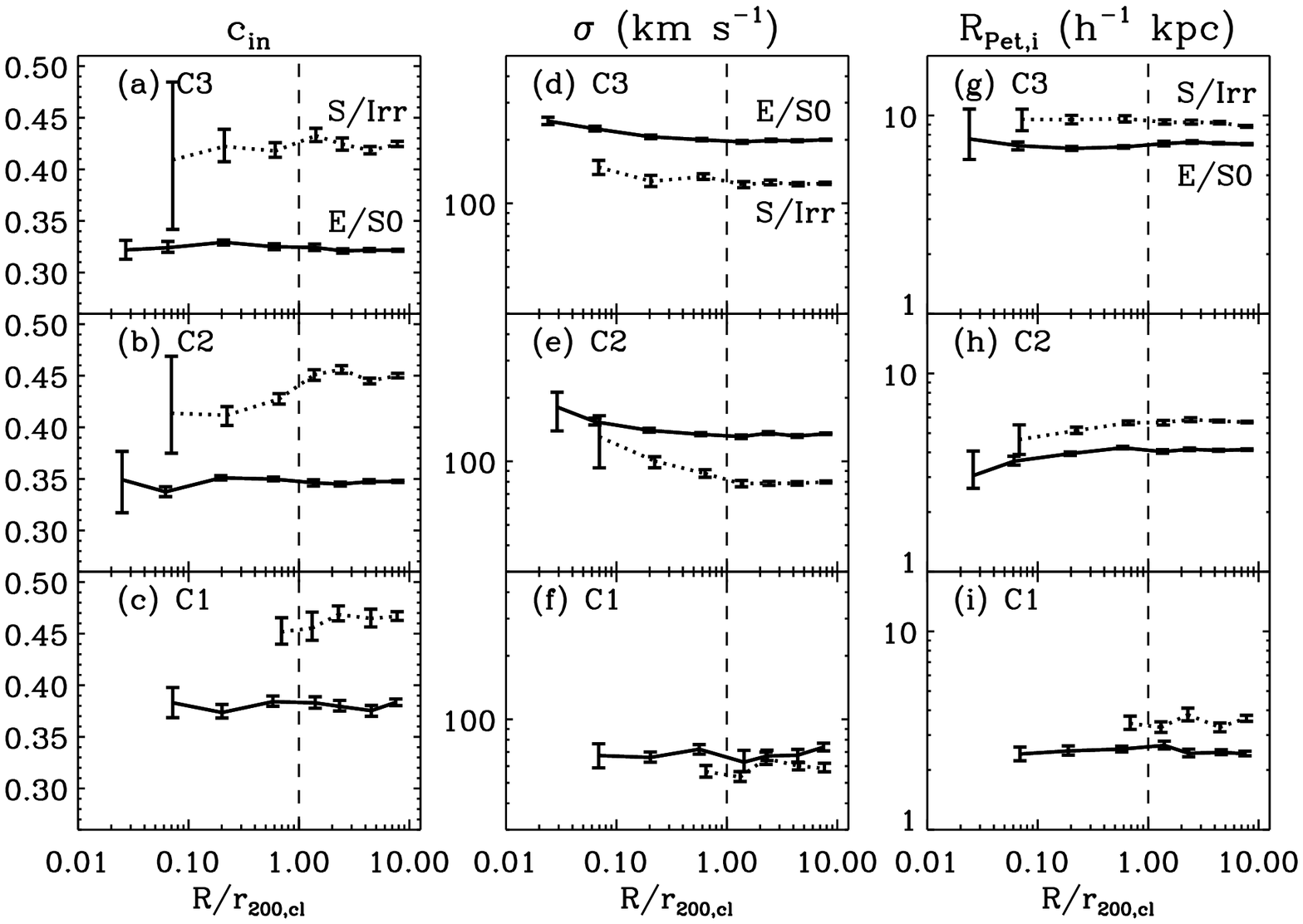}
\caption{
Same as Fig. \ref{fig-sf1d}, but for
  (left) $c_{\rm in}$, (middle) $\sigma$ (km s$^{-1}$), and (right) R$_{\rm Pet,i}$ ($h^{-1}$ kpc).
  Late-types with axis ratio of $b/a<0.6$ were included.
}\label{fig-struc1d}
\end{figure*}

The equivalent width of the hydrogen $H\alpha$ line is a measure of SFA. 
Figure \ref{fig-sf1d} shows the median $W(H\alpha)$
  as a function of $R$ for six different cases distinguished by
  morphology and luminosity of target galaxies.
The SFA of late-type galaxies decreases as $R$ decreases.
The dependence again seems to start at $\sim3r_{\rm 200,cl}$ for the galaxies
  in the C1 sample.
It is important to note that both $u-r$ color and $W(H\alpha)$ show
  only mild dependence on the clustercentric radius $R$,
  and it is actually the neighboring galaxies that mainly determine the color
  and SFA of galaxies within clusters as we can see in Figure \ref{fig-sf2d}.

The second column of Figure \ref{fig-sf2d} shows $W(H\alpha)$ in
  two-dimensional environmental parameter space. 
One can notice that the constant $W(H\alpha)$ contours look very similar to those
  of $u-r$ color.
Namely, the SFA of the L-e galaxies is mainly controlled by the neighbor galaxies.
On the other hand, the SFA of L-l galaxies shows dependence on $R$
  when the nearest neighbor separation is not too small ($R_n\ga0.3r_{\rm vir,n}$).
This is again probably because the neighbors other than the nearest one tend to
  be early types as $R$ decreases.
The two cases demonstrate that the SFA of the late-type cluster galaxies
  changes differently depending on the morphology of the neighbor galaxy
  even well inside the cluster virial radius.
If the hot cluster gas could directly quench the SFA of late-type galaxies,
  both L-l and L-e galaxies would show the same $R$-dependence of SFA
regardless of the morphology of the nearest neighbor.
Since this is not the case, one can conclude that it is after all
  the galaxy-galaxy hydrodynamic interaction that gives the biggest impacts on
  the color and SFA of cluster galaxies and that,
  contrary to intuition, the hot cluster gas is not the main actor
  quenching the star formation in late-type cluster galaxies at
  least down to $R\approx0.1r_{\rm 200,cl}$.
  We have repeated our analyses using the fainter galaxies with 
$-17.5\ge M_r > -19.0$, and arrived at the same conclusion.
In the case of early-type galaxies
the $W(H\alpha)$ parameter is very small and hardly changes. 

%The SFA of late-type galaxies drops significantly within the radius.
%Since the change is rather abrupt as was seen for the previous
%parameters, it is reasonable to associate this phenemenion with 
%the hydrodynamic origin. It can be learned that the cluster hot
%gas quenches the SFA of late-type galaxies significantly and 
%the effect becomes stronger as $R$ decreases.
%However, the closest neighbor jointly affects the SFA within
%the cluster virial radius when it is an early type.
%The effects of neighbor start to be dominant at $R_n \la 0.2 
%r_{200,nei}$ within the cluster virial radius.

Figure \ref{fig-sf1d} shows that the color gradient of galaxies depends only weakly on $R$. 
In all cases the central region of galaxies 
  becomes slightly bluer relative to the outskirts as $R$ decreases.
The dependence is stronger for fainter galaxies.

In the right column of Figure \ref{fig-sf2d} contours represent the distribution of the median $\Delta (g-i)$
  at each location of the $R$-$R_n$ space.
Unlike the $u-r$ color and $W(H\alpha)$, the color gradient $\Delta (g-i)$ of late-type galaxies
  depends on both $R$ and $R_n$ inside $r_{\rm 200,cl}$
  as can be seen from the slant contours.
We interpret this phenomenon as a result of galaxy-galaxy interaction.
When a late-type galaxy approaches a neighbor closer than the virial radius of the neighbor,
  its color gradient always increases (center becomes relatively bluer)
  regardless of neighbor's morphology as shown by \citet[see Figs. 6 and 7]{pc09}.
The color gradient of a late-type galaxy 
  can increase as it moves toward the cluster center because it becomes more likely to be
  affected by neighbor galaxies.
Since there is no neighbor morphology dependence of $\Delta (g-i)$
  when a late type approaches a neighbor,
  the $\Delta (g-i)$ contours are quite similar for L-e and L-l galaxies.
It is also possible that the center of late-type cluster galaxies becomes bluer
at smaller $R$ because of the tidal effects of the cluster potential \citep{mer84}.

\subsection{Structure Parameters}

%Figure 12 %%%%%%%%%%%%%%%%%%%%%%%%%%%%%%
\begin{figure*}
\center
\plotone{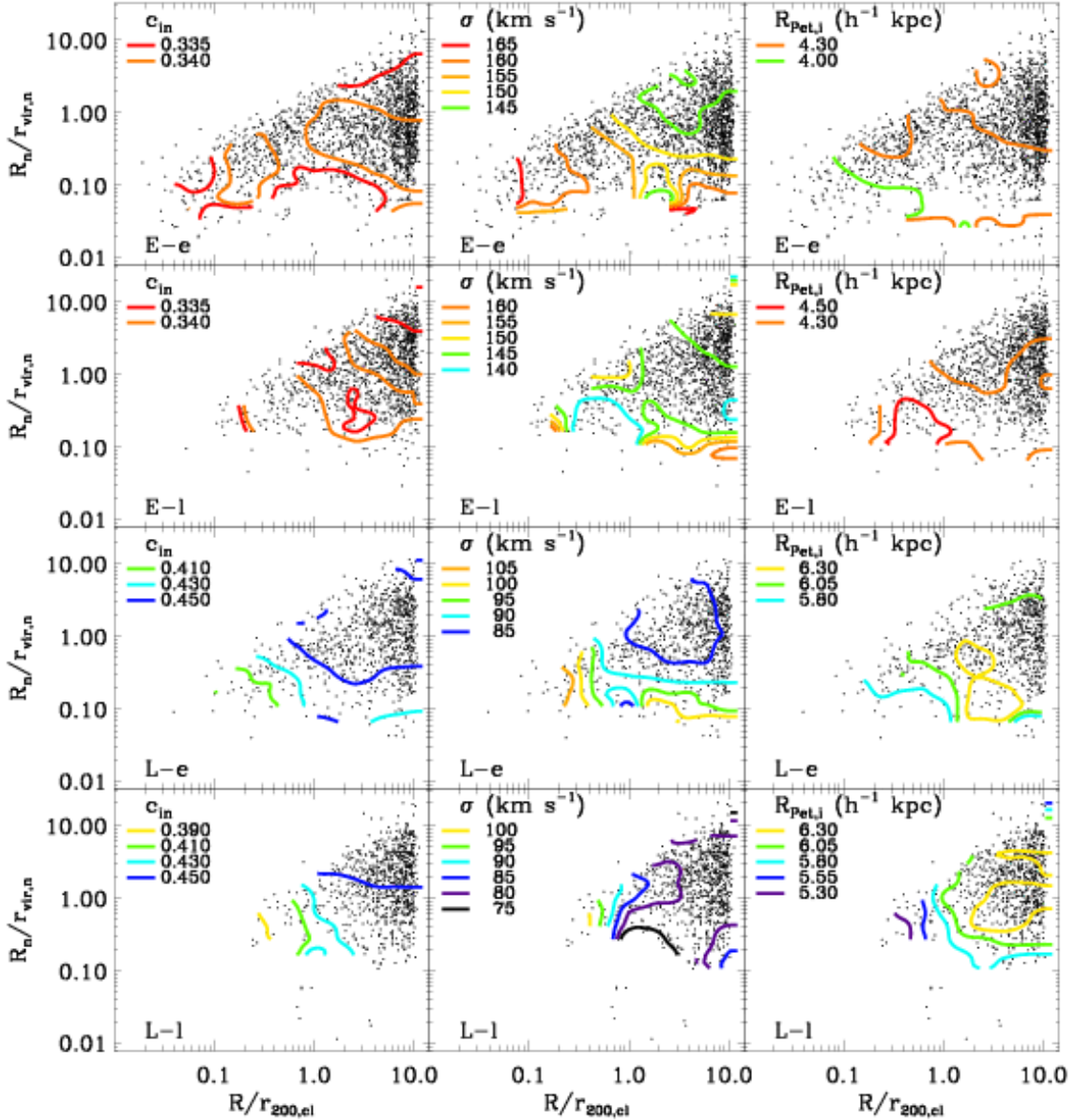}
\caption{
Same as Fig. \ref{fig-sf2d}, but for
  (left) $c_{\rm in}$, (middle) $\sigma$ (km s$^{-1}$), and (right) 
R$_{\rm Pet,i}$ ($h^{-1}$ kpc).
  Late-types with axis ratio of $b/a<0.6$ were included.}
\label{fig-struc2d}
\end{figure*}

The left column of Figure \ref{fig-struc1d} shows that concentration of intermediate and low
  luminosity late-type galaxies increases ($c_{in}$ decreases) as $R$ decreases below $R_{\rm cr}$.
But concentration of early types and high luminosity late types is nearly independent of $R$. 
As argued by \citet{pc09}, this can be attributed to less compact internal structure of late types
  which are more vulnerable to tidal effects than early types.

The first column of Figure \ref{fig-struc2d} shows the median $c_{in}$ contours 
  for the intermediate luminosity galaxies with $-19.5\ge M_r>-20.5$.
Again $c_{in}$ of early types hardly 
changes as $R$ or $R_n$ vary. On the other hand, $c_{in}$ of late types 
starts to decrease (galaxies become compacter) at $R\la  R_{\rm cr}=1\sim3 r_{\rm 200,cl}$. 
It is interesting to see that the structural parameter $c_{in}$ shows an
  abrupt change at the characteristic scale. 
A similar phenomenon can be found also for the central velocity dispersion. 

%It should be noted that inside $R=R_{\rm cr}$ the $c_in$ parameter depends not 
%only $R$ but also $R_n$ as indicated by skewed contour in the L-e and L-l 
%panels. It can be also noted that the contours in there panels are guite 
%similar to those in the corresponding panels for $u-r, W(H\alpha)$ and 
%$\delta(g-i)$ parameters, the indicators for star formation activity. 

The fact that the $R$-dependence of $c_{in}$ exists only when $R\la R_{\rm cr}$,
  gives an important clue for understanding the structural evolution of cluster galaxies.
The local galaxy number density cannot be the reason for the discontinuous $R$-dependence of $c_{in}$
  because the local density is a smooth function of $R$.
The discontinuity can appear due to repeated gravitational interactions of cluster member galaxies
  with cluster potential or with other galaxies as they make trapped orbital motions within
  the cluster virial radius.
It might also seem possible to explain the discontinuity by the hydrodynamic effects of hot cluster gas.
For example, the cold gas in the outer part of late-type galaxies can evaporate or be 
  stripped when they fall into the hot cluster gas clump or encounter 
  the hot halo gas of their neighbor galaxies. 
Then their outer part will become redder and dimmer. 
This may cause the color $u-r$ and color gradient $\Delta(g-i)$ increase and 
  inverse concentration index $c_{in}$ decreases as $R$ or $R_n$ decreases. 
However, it is difficult to explain why the central stellar velocity dispersion
  of late types should also increase shown in the second column of Figure \ref{fig-struc2d}
  by quenching the SFA in the outer part.
  
Figure \ref{fig-struc1d} (panels d, e, and f) shows that 
  the central velocity dispersion of galaxies in general increases 
  as $R$ decreases except for low luminosity galaxies. 
The trend is clear for intermediate luminosity galaxies, in particular.

The second column of Figure \ref{fig-struc2d} shows constant $\sigma$ contours of 
  intermediate luminosity ($-19.5\ge M_r>-20.5$) galaxies divided into 
  four different target-neighbor morphology cases. 
The central stellar velocity dispersion $\sigma$ 
  shows a characteristic behavior that 
  it depends only on $R_n$ at $R \ga R_{\rm cr}$ but starts to depend only on $R$ 
  when $R \la R_{\rm cr}$.
%As we pointed out above, it is difficult to explain the change in $\sigma$  
%  by hydrodynamic interactions.
%Therefore, the sudden dependence of $\sigma$ on $R$ inside $R_{\rm cr}$ should be
%  attribute to gravitational interaction of galaxies with neighbor galaxies or cluster potential.
Would the increase of $\sigma$ within the cluster virial radius be 
  due to the tidal interaction of galaxies with other galaxies or with the cluster potential? 
One can find a clue from Figure \ref{fig-struc1d} which shows that
  the $R$-dependence of $c_{in}$ and $\sigma$ is largest for
  intermediate luminosity galaxies.
This may be because the intermediate luminosity galaxies have suffered from the interaction
  with cluster potential more strongly than brighter or fainter ones.
Even though galaxy-galaxy encounters in general increase 
  $\sigma$ of late-type galaxies (see L-e cases at $R>r_{\rm 200,cl}$ and Fig. 8 of \citealt{pc09}),
  it is not true at $R<r_{\rm 200,cl}$ as can be seen in Figure \ref{fig-struc2d}.
The very bright massive galaxies at $0.1\la R/r_{\rm 200,cl} \la 1.0$ may be
  the galaxies falling into the cluster for the first time, and those
  which fell in long time ago are now sitting near the cluster center.
In both cases the effects of the cluster potential will be relatively small.
On the other hand, the small faint galaxies are likely to have large orbital radii,
  and can cross the cluster much fewer times than the massive ones.
This is supported by the fact that $R_{\rm cr}$ is larger for fainter galaxies 
  (see Fig. \ref{fig-fracdd} for example).
Therefore, it is expected that the intermediate luminosity galaxies
  are most susceptible to change in structure and kinematics
  through interactions with the cluster potential.

Our study of $c_{in}$ and $\sigma$ of cluster galaxies makes it clear that 
  late-type galaxies become more centrally concentrated and have higher 
  central velocity dispersion as they approach the cluster center.
Our results suggest that late-type galaxies become earlier in SFA
  through hydrodynamic interactions between galaxies and
  also in morphology and kinematics through tidal interactions
  as they approach the cluster center. 
Paucity of late-type galaxies near cluster center is a result of combined 
  effects of gravitational and hydrodynamic interaction with
  the cluster and the nearest neighbor galaxies. 

%\subsection{Size}
 
We use the Petrosian radius in the $i$-band image as a measure of galaxy size. 
Figure \ref{fig-struc1d} (panels g, h, and i) shows that
  the intermediate luminosity galaxies show 
  very slight decrease of $R_{\rm Pet}$ as $R$ decrease.
The third column of Figure \ref{fig-struc2d} shows again that 
  there is a weak clustercentric radius dependence of $R_{\rm Pet}$ at $R \la R_{\rm cr}$
  for late-type galaxies.
The decrease of $R_{\rm Pet}$ does not seem to be due to the blending
  of galaxy images in high density regions like clusters because it suddenly
  occurs at the physically meaningful scale ($R_{\rm cr}$).
The degree of blending of galaxy images should be a smooth function of $R$.

The size of late-type galaxies does not change significantly by interactions
  with other galaxies as shown in Figure 8 of \citet{park08}.
Therefore, the decrease of $R_{\rm Pet}$ of late types at $R<R_{\rm cr}$ shown in 
  Figures \ref{fig-struc1d} and \ref{fig-struc2d} should be attributed to the cluster.
Late-type galaxies can appear smaller when their outer part becomes dimmer
  as they orbit within a cluster and get quenched in star formation in disks.

\section{DISCUSSION}

Many mechanisms have been proposed to explain the MRR or MDR in clusters. 
They are divided into two categories : mechanisms relying on gravitational and hydrodynamic processes 
  (see \citealt{bg06} for a review).
The mechanisms based on gravitational interactions include galaxy-galaxy tidal interaction, 
  galaxy-cluster potential tidal interaction, and interaction with the 
  general tidal background (harassment). 
Those based on hydrodynamic interactions comprise ram pressure stripping, 
  viscous stripping, or thermal evaporation of cold gas of late-type galaxies 
  by hot cluster gas, and removal of hot halo gas reservoir causing stopping of gas 
  supply and quenching of star formation (strangulation). 
 We will examine each of these mechanisms against our findings.

 1. Galaxy-galaxy tidal interaction

Tidal interaction between galaxies 
  can be efficient at removing or consuming the cold gas in galaxies,
  and tend to transform late types into early types \citep{sb51,ric76,fs81,icke85,mer83}.
However, this mechanism is usually excluded as a process responsible for 
  the MRR or MDR because the orbital velocity of cluster galaxies are 
  very high and the tidal energy deposit during the short encounters 
  is too small to significantly affect galaxy structure \citep{mer84,bv90}. 
% We exclude all tidal interaction-based mechanisms due to the fact 
%that gravity has no characteristic scale while we found there exists 
%a characteristic radius $R_c \approx 2\sim 3 r_{\rm 200,cl}$ where 
%galaxy morphology and star formation activity parameters suddenly 
%start to depend on the clustercentric radius $R$. 

At $R>R_{\rm cr}$, the structural parameter of late-type galaxies depend only on
  the environment determined by the neighbor galaxies.
But $R<R_{\rm cr}$ they strongly depend on $R$, but only weakly or negligibly on $R_n$ 
  (see Fig. \ref{fig-struc2d}).
This tells that the galaxy-galaxy tidal interactions do not instantly change the structure
  of cluster member galaxies much.
The parameters $c_{in}$ and $R_{\rm Pet}$ do show a weak dependence on $R_n$ for L-e galaxies
  within $R=R_{\rm cr}$.
But this may have also been resulted by hydrodynamic interactions with their early-type neighbor.
Even though we exclude the significance of the tidal effects of galaxy-galaxy interactions
  on the internal structure and kinematics of cluster galaxies, 
  it should be emphasized that the hydrodynamic effects of 
  galaxy-galaxy interactions can give significant impacts on their SFA.

 2. Galaxy-cluster potential interaction

Individual galaxies can suffer from the tidal force of the overall cluster mass
 \citep{mer84,mil86,bv90,mw00,gne03}.
This mechanism is usually excluded in explaining the MRR or MDR because the time 
  scale required for morphology transformation, being several cluster crossing time, 
  is rather too long.
The tidal force will be stronger toward the cluster center,
  and the strong dependence of $c_{in}$ and $\sigma$ on $R$ within $R=R_{\rm cr}$
  can easily explained by this mechanism.
The internal structure of the intermediate mass galaxies, in particular,
  seems to have been significantly affected by this mechanism.
But the tidal interaction with cluster potential may not be sufficient for
  morphology transformation of the very bright or faint cluster galaxies.

% We exclude this mechanism because it cannot explain the existence of 
%the characteristic scale $R_c$. Even though the tidal effects of galaxy-cluster 
%interaction on cluster galaxy properties are excluded, the hydrodynamic effects 
%of galaxy-cluster interaction seem important. 

3. Harassment

Galaxies can be also perturbed by the tidal force background from numerous 
  distant encounters with other cluster members \citep{moo96,moo98,moo99}. 
Even though individual encounters are too short for the tidal force to activate
  structural changes, galaxy structure may significantly 
change through such numerous weak encounters.
The sudden transition of the structure parameters can be explained by repeated passage
  of member galaxies within the cluster.
So both interaction with cluster potential and harassment may account for
  the $R$-dependence of morphology and structure parameters of intermediate
  mass galaxies, in particular, whose cluster crossing time is reasonably short
but the path length is still long.
%We again exclude harassment 
%as a working mechanism responsible for the MRR/MDR because it cannot have 
%a characteristic scale where it suddenly turns on. 

%Figure 13 %%%%%%%%%%%%%%%%%%%%%%%%%%%%%%
\begin{figure*}
\center
%\plotone{f16.eps}
\includegraphics[scale=0.6]{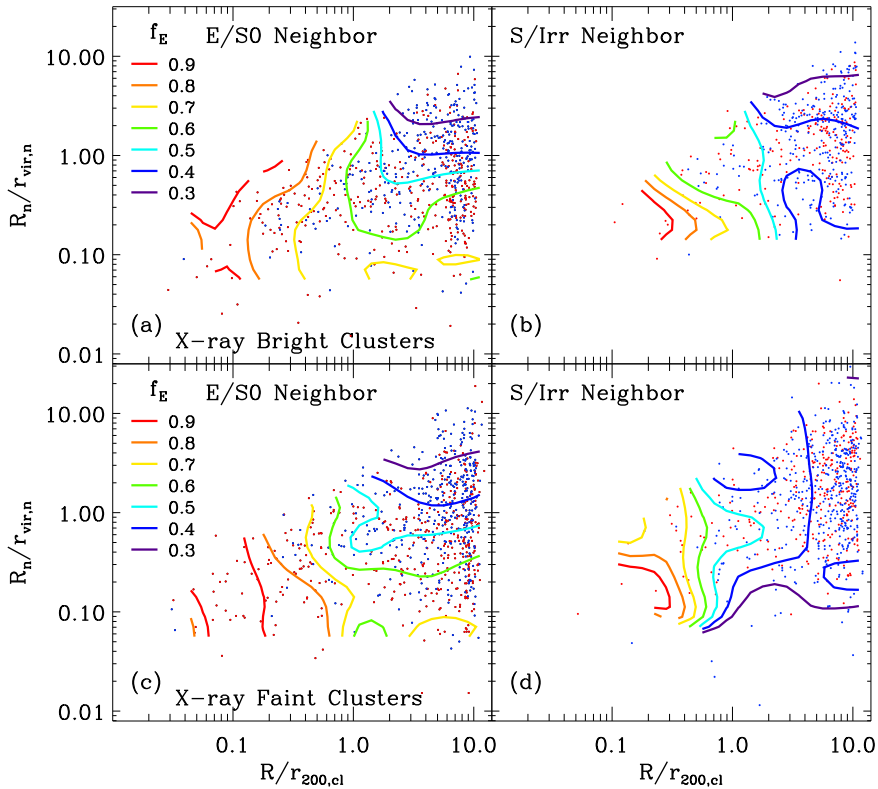}
\caption{
Same as Fig. \ref{fig-frac2d}, but for
  (a and b) X-ray bright clusters ($L_X\geq 2\times10^{44}$ erg s$^{-1}$) and
  (c and d) X-ray faint clusters ($0.6\times10^{44}\leq L_X\leq2\times10^{44}$ erg s$^{-1}$).
   X-ray luminosities are from \citet{boh00,boh04}.
}\label{fig-fracxray}
\end{figure*}

4. Galaxy-galaxy hydrodynamic interaction

The effects of hydrodynamic interaction between galaxies on cluster galaxy 
  properties have been found important in this work. 
The galaxy-galaxy interaction turns out to be the major mechanism 
  for quenching SFA of late-type galaxies within clusters.
As can be seen in the scatter plot of galaxies in the $R$-$R_n$ plane,
  distance to the neighbors monotonically decreases
  as $R$ decreases and galaxies become always located
  within neighbor galaxy's virial radius when $R\la R_{\rm 200,cl}/3$.
In this situation galaxies are constantly undergoing hydrodynamic effects and 
  mass exchange with neighbors. 
We argued above that the vertical contours for $c_{in}$ and $\sigma$ can be explained by
  the gravitational tidal force from cluster itself.
However, the nearly horizontal contours for the $u-r$ and $W(H\alpha)$ parameters can be
  only explained by hydrodynamic effects of neighbor galaxies.
It should be emphasized that the high-speed orbital motion of cluster galaxies weakens
  the importance of galaxy-galaxy tidal interactions, but can make the galaxy-galaxy
  hydrodynamic interaction very important.

It was also found that bright galaxies are more isolated from influential neighbors
  than relatively fainter ones 
  in clusters, which has been interpreted as due to merger-driven luminosity transformation.
  The fact that early-type galaxies show such trend more strongly
is consistent with this interpretation.
Relative paucity of bright galaxies near cluster center (except for the BCGs) seems 
  to indicate that the frequency of galaxy merger decreases as the clustercentric 
  radius decreases. 
%Even though only a small fraction of cluster galaxies  have neighbors within 
%$R_n=0.1 r_{200,nei}$, 
%the frequence of such close encounters increases and more and more galaxies 
%most have been experienced hydrodynamic interaction with other galaxies 
%during the lifetime of cluster as the clustercentric radius decreases.  
%However, this process may not result in the characteristic scale for galaxy 
%properties if it is statistically a function only of cluster galaxy number density. 
%One possibility for this process to yield such a scale is the maximum radius 
%at which typical clusers galaxies and reach that maximum radius where 
%they must back to the center. The maximum radius must be close to the 
%cluster virial radius.
% Therefore, it is possible for the galaxy-galaxy hydrodynmic interaction
% to account fro the MRR or MDR in clusters. 

5. Ram Pressure Stripping 

Many rich clusters like the Abell clusters we are analyzing, are holding hot X-ray emitting gas. 
A late-type galaxy falling into such a hot gas tank 
  can strip off its cold gas and terminate SFA \citep{gg72,qui00}. 
The hot gas is trapped roughly within the virial radius of the cluster. 
Therefore, any mechanism relying on hydrodynamic interaction with the hot 
  cluster gas will have a characteristic scale of onset of cluster influence 
  near the cluster virial radius.   

A major problem with the ram pressure stripping and other hydrodynamic 
  processes in accounting for the morphology segregation, is that these 
  mechanisms can only reduce the SFA and cannot change galaxy structure 
while the observations require the fraction of bulge-dominated galaxies 
  to increase significantly as the clustercentric radius decreases. 
Therefore, even though this mechanism can explain the radial variation 
  of the fraction of galaxies that are red and inactive, it can not explain the radial 
  variation of concentration or central velocity dispersion.
It is also pointed out that only loosely bound thin clouds can be swept
  out of the galaxy and molecular clouds are mostly unaffected by this interaction \citep{qui00}.
Furthermore, we have shown in Figure \ref{fig-sf2d} that 
  it is actually the galaxy-galaxy interaction that mainly controls
  the color and SFA of the cluster late-type galaxies rather than
  the hot cluster gas.

6. Viscous stripping / Thermal evaporation

The cold gas in a late-type galaxy moving through the hot intracluster gas can be 
  stripped off by momentum transfer between the cold disk gas and hot cluster gas \citep{nul82}, 
  or be evaporated by the thermal conduction \citep{cs77}. 
 These mechanisms can be activated when a galaxy fall into the hot cluster 
gas pool, and can give rise to a characteristic feature in the radial variation 
of galaxy SFA as observed in this work. 
 However, they share the same problem the ram-pressure stripping mechanism 
has in that they can not modify galaxy structure.

Furthermore, the hot cluster gas cannot explain the environmental dependence of color
  and SFA of late-type galaxies either.
If the hot cluster gas predominantly controls the color and SFA of cluster late types,
  the contours in Figure \ref{fig-sf2d}
should be vertical independently of $R_n$, which is not true.

7. Strangulation

 According to the current understanding of galaxy evolution, spiral galaxies 
maintain SFA by accreting gas from their hot gas reservoir. 
It has been proposed that a spiral galaxy fell into the hot intracluster gas can 
  lose its hot halo gas and stop supplying gas into its disk \citep{lar80,bek02}. 
Then the galaxy just consume its already existing cold gas to form stars,
  and the SFA halts  after some time when all cold gas is consumed. 
Compared to the ram pressure stripping, 
  this mechanism will quench the SFA in late-type galaxies in a delayed action. 
%Since it takes some time for the late-type galaxies to consume the cold 
%gas they have been carrying. The delay time scale is estimated to be (~ ).
% A result of this time delay in quenching star formation activity is that it is not 
%expected for this mechanism to cause a characteristic scale in the 
%radial variation of galaxy properties. 
As being a pure hydrodynamic process, it also has the problem that the structural 
  parameters of galaxies can not be changed.
This mechanism also predicts the contours to be vertical (dominance of cluster influence)
  for the early-type fraction, color, and SFA parameters in Figures \ref{fig-frac2d} and \ref{fig-sf2d},
  which is not supported by observations.

To inspect galaxy properties for their dependence on hot cluster gas in more details,
  we studied the early-type fraction in the $R_n$-$R$ space separately for 
  X-ray bright ($L_x \geq 2\times 10^{44}$ erg s$^{-1}$)
  and faint ($0.6\times10^{44} \leq L_x \leq 2\times 10^{44}$ erg s$^{-1}$) Abell clusters.
Figure \ref{fig-fracxray} shows that the magnitude of $f_E$ is not particularly higher for
  X-ray bright clusters.
This can be another evidence against the idea that the hot intracluster gas plays the main role
  in morphology transformation of late-type galaxies in massive clusters.
  
To summarize, the interaction responsible for the MRR/MDR in clusters can be
  either galaxy-galaxy or galaxy-cluster interactions from the point of view of the actor, 
  or either gravitational (mass-mass) or hydrodynamic 
  (gas-gas) interactions in terms of physical process.
%We argued for the hydrodynamic 
%galaxy-cluster interaction because it can explain the characteristic 
%scale of $R_c=2~3r_{\rm 200,cl}$ in the radial variation of galaxy SFA. We also 
%support for the hydrodynamic galaxy-galaxy interaction since our 
%work show that his process is acting within clusters. 
Figure \ref{fig-struc2d} suggests that the galaxy-cluster tidal interactions are responsible for
  structural and kinematic changes of cluster late-type galaxies toward early morphological type.
Late types seem to become more spheroidal through such interactions.
Figure \ref{fig-sf2d} indicates that the hydrodynamic interactions 
  with early-type neighbors are responsible for reddening and SFA quenching 
  of cluster late-type galaxies.
Therefore, the morphology transformation of late-type galaxies seems to take place
  in clusters through galaxy-galaxy hydrodynamic interactions and galaxy-cluster/galaxy-galaxy
  gravitational interactions.
However, it is reasonable to accept that all above mechanisms are contributing 
  to the MRR/MDR in clusters to some extent and a particular process cannot 
  fully account for all observational aspects.

% Based on existence of the characteristic scale $R_{\rm cr}$, we propose the 
%galaxy-cluster hydrodynamic interactions like ram pressure stripping, 
%viscous stripping, and thermal evaporation are making an important 
%role in the MRR/MDR. However, we have found clear evidence that 
%galaxy-galaxy hydrodynamic interaction gives a major control in 
%determining properties of clse pairs of galaxies in clusters. 
%Considering the fact that galaxy-galaxy hydrodynamic interaction 
%is the main cause for change in of cluster galaxies, ..... 

\section{Conclusions}

%Morphology transformation due to tidal interaction with the cluster potential and
%harassment by the general tidal field in clusters cannot explain
%the $f_E-R$ relation when $R< r_{\rm 200,cl}$ and $R_n \ga 0.1 r_{vir,n}$
%because there is no reason for these mechanisms
%to be suddenly effective within $R = r_{\rm 200,cl}$. Tidal interactions between
%individual galaxies predict that interactions with an early-type neighbor
%or a late-type neighbor give rise to morphology transformation in the same 
%direction. However, one can see in Figure * that the direction of 
%morphology transformation changes depending on the morphological type
%of the nearest neighbor galaxy.
%Therefore, the sensitivity of $f_E$ to $R_n$ at $R< r_{\rm 200,cl}$ and $R_n\la 0.1
%r_{vir,n}$ cannot be due to the gravitational interactions among galaxies. 
%Since the mechanism is not effective at this short neighbor separations,
%it must not be effective at larger separations ($R_n \ga 0.1 r_{vir,n}$) 
%as well.

We have studied the dependence of various galaxy properties on the clustercentric
  radius and the environment attributed to the nearest neighbor galaxy
  using the SDSS galaxies associated with the Abell galaxy clusters.
Our major findings are as follows.

1. There exists a characteristic scale where the galaxy properties such as 
morphology, color gradient, and structural parameters suddenly start
to depend on the clustercentric radius at fixed nearest neighbor environment.

2. The characteristic scale depends on galaxy luminosity; the faint galaxies
with $-17.0\ge M_r > -19.0$ has the scale at $\sim 3 r_{\rm 200,cl}$ while
the scale is $\sim r_{\rm 200,cl}$ for the brighter galaxies with
$-20.5\ge M_r > -22.5$.

3. The hydrodynamic interactions with nearby early-type galaxies seem
to be the main drive to quenching star formation activity of late-type
galaxies in clusters. We do not find evidence that the hot cluster
gas is the main drive down to the clustercentric radius of $\sim 0.1
r_{\rm 200,cl}$.

4. The interaction with the cluster potential and harassment are the 
viable mechanisms that can account for the clustercentric radius
dependence of the structural and internal kinematics parameters.

Existence of the characteristic scale means that the local galaxy
  number density is not responsible for the MDR
  in cluster because the local density is a smooth function of the 
  clustercentric radius and has no discontinuity in general. 
The MDR appears working in clusters and also in the field because of the statistical
  correlation between the local density and the nearest neighbor distance.
What is really working in clusters is the morphology-clustercentric
  radius-neighbor environment relation, where the neighbor environment
  means both neighbor morphology and the mass density attributed to the neighbor.

According to the third and fourth conclusions, the morphology 
  transformation of the late-type galaxies in clusters is not due to
  a single mechanism and has been taken place through both
  hydrodynamic and gravitational processes.

\acknowledgments
We would like to thank the anonymous referee for useful comments.
The authors acknowledge the support of the Korea Science and Engineering
Foundation (KOSEF) through the Astrophysical Research Center for the
Structure and Evolution of the Cosmos (ARCSEC).
% and through the grant R01-2004-000-10520-0. 
%JRG is supported by NSF GRANT AST 04-06713.

Funding for the SDSS and SDSS-II has been provided by the Alfred P. Sloan 
Foundation, the Participating Institutions, the National Science 
Foundation, the U.S. Department of Energy, the National Aeronautics and 
Space Administration, the Japanese Monbukagakusho, the Max Planck 
Society, and the Higher Education Funding Council for England. 
The SDSS Web Site is http://www.sdss.org/.

The SDSS is managed by the Astrophysical Research Consortium for the 
Participating Institutions. The Participating Institutions are the 
American Museum of Natural History, Astrophysical Institute Potsdam, 
University of Basel, Cambridge University, Case Western Reserve University, 
University of Chicago, Drexel University, Fermilab, the Institute for 
Advanced Study, the Japan Participation Group, Johns Hopkins University, 
the Joint Institute for Nuclear Astrophysics, the Kavli Institute for 
Particle Astrophysics and Cosmology, the Korean Scientist Group, the 
Chinese Academy of Sciences (LAMOST), Los Alamos National Laboratory, 
the Max-Planck-Institute for Astronomy (MPIA), the Max-Planck-Institute 
for Astrophysics (MPA), New Mexico State University, Ohio State University, 
University of Pittsburgh, University of Portsmouth, Princeton University,
the United States Naval Observatory, and the University of Washington. 

This research has made use of the NASA/IPAC Extragalactic Database (NED) 
which is operated by the Jet Propulsion Laboratory, California Institute of Technology, 
under contract with the National Aeronautics and Space Administration.

\appendix

We calculate the cluster velocity dispersion by using
the cluster member galaxies after excluding the interlopers.
To find the interlopers among the cluster galaxies,
  we compute $\delta$ for each galaxy, which indicates
  the local deviations from the systemic velocity ($v_{\rm sys}$) and 
  dispersion ($\sigma_{\rm cl,all}$) of the entire cluster \citep{ds88}.
%We compute $\delta$ for each galaxy, which indicates
%  the local deviations from the systemic velocity ($v_{\rm sys}$) and 
%  dispersion ($\sigma_{\rm cl,all}$) of the entire cluster \citep{ds88}.
It is defined by
\begin{equation}
\delta^2 =  \frac{N_{nn}}{\sigma_{\rm cl,all}^2} \left[ (v_{\rm local}-v_{\rm sys})^2 +
(\sigma_{\rm local}-\sigma_{\rm cl,all})^2 \right],
\end{equation}
where $N_{nn}$ is the number of the nearest galaxies that defines the local environment, taken
  to be ${N_{\rm gal}}^{1/2}$ in this study.
${N_{\rm gal}}$ is the total number of member galaxies in the cluster.
The nearest galaxies are those located closest to the galaxy on the sky.
$v_{\rm local}$ and $\sigma_{\rm local}$ are
  systematic velocity and its dispersion estimated from $N_{nn}$ nearest galaxies, respectively.
We use the galaxies with $\delta\leq3.0$ to compute the cluster velocity dispersion.


\begin{thebibliography}{}

\bibitem[Abell et al.(1989)]{aco89} Abell G. O., Corwin H. G., \& Olowin R. P., 1989, \apjs, 70, 1

\bibitem[Adelman-McCarthy et al.(2008)]{ade08} Adelman-McCarthy, J.~K., et al.\ 2008, \apjs, 175, 297 

\bibitem[Ann et al.(2008)]{ann08} Ann, H.~B., Park, C., 
\& Choi, Y.-Y.\ 2008, \mnras, 389, 86 

%\bibitem[Bamford et al.(2009)]{bam09} Bamford, S.~P., et al.\ 
%2009, \mnras, 393, 1324 

\bibitem[Bekki et al.(2002)]{bek02} Bekki, K., Couch, W.~J., 
\& Shioya, Y.\ 2002, \apj, 577, 651 

\bibitem[Blanton et al.(2003)]{bla03} Blanton, M.~R., Lin,
H., Lupton, R.~H., Maley, F.~M., Young, N., Zehavi, I., \&
Loveday, J.\ 2003, \aj, 125, 2276

\bibitem[Blanton et al.(2005)]{bla05} Blanton, M.~R., Eisenstein, D., Hogg, D.~W., Schlegel, D.~J.,
\& Brinkmann, J. 2005, \aj, 129, 2562

\bibitem[B{\"o}hringer et al.(2000)]{boh00} B{\"o}hringer, H., et al.\ 2000, \apjs, 129, 435 

\bibitem[B{\"o}hringer et al.(2004)]{boh04} B{\"o}hringer, H., et al.\ 2004, \aap, 425, 367 

\bibitem[Boselli \& Gavazzi(2006)]{bg06} Boselli, A., \& Gavazzi, G.\ 2006, \pasp, 118, 517 

\bibitem[Bruzual \& Charlot(2003)]{bc03} Bruzual, G., \& Charlot, S.\ 2003, \mnras, 344, 1000

\bibitem[Byrd \& Valtonen(1990)]{bv90} Byrd, G., \& Valtonen, M.\ 1990, \apj, 350, 89 

\bibitem[Carlberg et al.(1997)]{car97} 
Carlberg, R.~G., Yee, H.~K.~C., \& Ellingson, E.\ 1997, \apj, 478, 462

\bibitem[Castander et al.(2001)]{cas01} Castander, F.~J., et al.\ 2001, \aj, 121, 2331

\bibitem[Choi et al.(2007)]{choi07} Choi, Y.-Y., Park, C., \& Vogeley, M.~S.\ 2007, \apj, 658, 884

\bibitem[Cowie \& Songaila(1977)]{cs77} Cowie, L.~L., \& Songaila, A.\ 1977, \nat, 266, 501 

\bibitem[Christlein \& Zabludoff(2005)]{cz05} Christlein, D., \& Zabludoff, A.~I.\ 2005, \apj, 621, 201 

\bibitem[Dom{\'{\i}}nguez et al.(2001)]{dml01} 
Dom{\'{\i}}nguez, M., Muriel, H., \& Lambas, D.~G.\ 2001, \aj, 121, 1266 

\bibitem[Dressler(1980)]{dre80} Dressler, A.\ 1980, \apj, 236, 351 

\bibitem[Dressler \& Shectman(1988)]{ds88} Dressler A., \& Shectman S. A., 1988, \aj, 95, 985

\bibitem[Dressler et al.(1997)]{dre97} Dressler, A., et al.\ 1997, \apj, 490, 577 

\bibitem[Fadda et al.(1996)]{fadda96}
Fadda D., Girardi M., Giuricin G., Mardirossian F., \& Mezzetti M., ApJ, 473, 670

\bibitem[Farouki \& Shapiro(1981)]{fs81} Farouki, R., \& Shapiro, S.~L.\ 1981, \apj, 243, 32 

%\bibitem[Fujita(1998)]{fuj98} Fujita, Y.\ 1998, \apj, 509, 587 

\bibitem[Fukugita et al.(1996)]{fuk96} Fukugita, M.,
Ichikawa, T., Gunn, J.~E., Doi, M., Shimasaku, K., \& Schneider,
D.~P.\ 1996, \aj, 111, 1748

\bibitem[Gnedin(2003)]{gne03} Gnedin, O.~Y.\ 2003, \apj, 589, 752 

\bibitem[Goto et al.(2003)]{goto03} Goto, T., Yamauchi, C., 
Fujita, Y., Okamura, S., Sekiguchi, M., Smail, I., Bernardi, M., 
\& Gomez, P.~L.\ 2003, \mnras, 346, 601 

\bibitem[Gott \& Rees(1975)]{gr75} Gott, J.~R., \& Rees, M.~J.\ 1975, \aap, 45, 365 

\bibitem[Gunn \& Gott(1972)]{gg72} Gunn, J.~E., \& Gott, J.~R.\ 1972, \apj, 176, 1 

\bibitem[Gunn et al.(1998)]{gunn98} Gunn, J.~E., et al.\ 1998, \aj, 116, 3040

\bibitem[Gunn et al.(2006)]{gunn06} Gunn, J.~E., et al.\ 2006, \aj, 131, 2332

\bibitem[Hogg et al.(2001)]{hogg01} Hogg, D.~W., Finkbeiner,
D.~P., Schlegel, D.~J., \& Gunn, J.~E.\ 2001, \aj, 122, 2129

\bibitem[Hwang \& Lee(2007)]{hl07} Hwang, H.~S., \& Lee, M.~G.\ 2007, \apj, 662, 236

\bibitem[Hwang \& Lee(2008)]{hl08} Hwang, H.~S., \& Lee, M.~G.\ 2008, \apj, 676, 218

\bibitem[Hwang \& Park(2008)]{hp08} Hwang, H.~S., \& Park, C.\ 2008, \apj, submitted

\bibitem[Hubble \& Humason(1931)]{hh31} Hubble, E., \& Humason, M.~L.\ 1931, \apj, 74, 43 

\bibitem[Icke(1985)]{icke85} Icke, V.\ 1985, \aap, 144, 115 

\bibitem[Ivezi{\'c} et al.(2004)]{ive04} Ivezi{\'c}, {\v Z}.,
et al.\ 2004, Astronomische Nachrichten, 325, 583

\bibitem[Larson et al.(1980)]{lar80} Larson, R.~B., Tinsley, 
B.~M., \& Caldwell, C.~N.\ 1980, \apj, 237, 692 

\bibitem[Lupton et al.(2002)]{lup02} Lupton, R.~H., Ivezic,
Z., Gunn, J.~E., Knapp, G., Strauss, M.~A., \& Yasuda, N.\ 2002,
\procspie, 4836, 350

\bibitem[Merritt(1983)]{mer83} Merritt, D.\ 1983, \apj, 264, 24 

\bibitem[Merritt(1984)]{mer84} Merritt, D.\ 1984, \apj, 276, 26 

%\bibitem[Mihos(2004)]{mih04} Mihos, J.~C.\ 2004, Clusters of 
%Galaxies: Probes of Cosmological Structure and Galaxy Evolution, 277 

\bibitem[Miller(1986)]{mil86} Miller, R.~H.\ 1986, \aap, 167, 41 

\bibitem[Moore et al.(1996)]{moo96} 
Moore, B., Katz, N., Lake, G., Dressler, A., \& Oemler, A.\ 1996, \nat, 379, 613 

\bibitem[Moore et al.(1998)]{moo98} 
Moore, B., Lake, G., \& Katz, N.\ 1998, \apj, 495, 139 

\bibitem[Moore et al.(1999)]{moo99} 
Moore, B., Lake, G., Quinn, T., \& Stadel, J.\ 1999, \mnras, 304, 465 

\bibitem[Moss \& Whittle(2000)]{mw00} Moss, C., \& Whittle, M.\ 2000, \mnras, 317, 667 

\bibitem[Nulsen(1982)]{nul82} Nulsen, P.~E.~J.\ 1982, \mnras, 198, 1007 

\bibitem[Park \& Choi(2005)]{pc05} Park, C., \& Choi, Y.-Y.\ 2005, \apj, 635, L29

\bibitem[Park \& Choi(2009)]{pc09} Park, C., \& Choi, Y.-Y.\ 2009, \apj, 691, 1828

\bibitem[Park et al.(2007)]{park07} Park, C., Choi, Y.-Y., 
Vogeley, M.~S., Gott, J.~R., \& Blanton, M.~R.\ 2007, \apj, 658, 898 

\bibitem[Park et al.(2008)]{park08} Park, C., Gott, J.~R., \& Choi, Y.-Y.\ 2008, \apj, 674, 784 

\bibitem[Peebles(1993)]{pee93} Peebles, P.~J.~E.\ 1993,
Princeton Series in Physics, Princeton, NJ: Princeton University Press

\bibitem[Pier et al.(2003)]{pier03} Pier, J.~R., Munn, J.~A.,
Hindsley, R.~B., Hennessy, G.~S., Kent, S.~M., Lupton, R.~H., \&
Ivezi{\'c}, {\v Z}.\ 2003, \aj, 125, 1559

\bibitem[Poggianti et al.(2008)]{pog08} Poggianti, B.~M., et al.\ 2008, \apj, 684, 888 

\bibitem[Postman et al.(2005)]{post05} Postman, M., et al.\ 2005, \apj, 623, 721 

\bibitem[Quilis et al.(2000)]{qui00} Quilis, V., Moore, B., 
\& Bower, R.\ 2000, Science, 288, 1617 

\bibitem[Quintero et al.(2006)]{qui06} Quintero, A.~D., 
Berlind, A.~A., Blanton, M., \& Hogg, D.~W.\ 2006, \apj, submitted (astro-ph/0611361)

\bibitem[Richstone(1976)]{ric76} Richstone, D.~O.\ 1976, \apj, 204, 642 

\bibitem[Schlegel et al.(1998)]{sch98} Schlegel, D.~J., 
Finkbeiner, D.~P., \& Davis, M.\ 1998, \apj, 500, 525 

\bibitem[Skibba et al.(2008)]{ski08} 
Skibba R.~A., et al.\ 2008, \mnras, submitted (astro-ph/0811.3970)

\bibitem[Smith et al.(2002)]{smi02} Smith, J.~A., et al.\
2002, \aj, 123, 2121

\bibitem[Smith et al.(2005)]{smi05} Smith, G.~P., Treu, T., 
Ellis, R.~S., Moran, S.~M., \& Dressler, A.\ 2005, \apj, 620, 78 

\bibitem[Spitzer \& Baade(1951)]{sb51} Spitzer, L.~J., \& Baade, W.\ 1951, \apj, 113, 413 

\bibitem[Stoughton et al.(2002)]{sto02} Stoughton, C., et al.\ 2002, \aj, 123, 485

\bibitem[Tegmark et al.(2004)]{teg04} Tegmark, M., et al.\ 
2004, \apj, 606, 702 

\bibitem[Thomas \& Katgert(2006)]{tk06} Thomas, T., \& Katgert, P.\ 2006, \aap, 446, 31 

\bibitem[Tremonti et al.(2004)]{tre04} Tremonti, C.~A., et al.\ 2004, \apj, 613, 898

\bibitem[Treu et al.(2003)]{treu03} Treu, T., Ellis, R.~S., 
Kneib, J.-P., Dressler, A., Smail, I., Czoske, O., Oemler, A., 
\& Natarajan, P.\ 2003, \apj, 591, 53 

\bibitem[Tucker et al.(2006)]{tuc06} Tucker, D.~L., et al.\
2006, Astronomische Nachrichten, 327, 821

\bibitem[Uomoto et al.(1999)]{uom99} Uomoto, A., et al.\
1999, Bulletin of the American Astronomical Society, 31, 1501

\bibitem[Weinmann et al.(2006)]{wei06} Weinmann, S.~M., van 
den Bosch, F.~C., Yang, X., \& Mo, H.~J.\ 2006, \mnras, 366, 2 

\bibitem[Whitmore et al.(1993)]{whi93} Whitmore, B.~C., 
Gilmore, D.~M., \& Jones, C.\ 1993, \apj, 407, 489 

\bibitem[York et al.(2000)]{york00} York, D.~G., et al.\ 2000, \aj, 120, 1579 

\end{thebibliography}
\end{document}